\documentclass[12pt]{article}
\usepackage[T1]{fontenc}
\usepackage{amssymb}
\usepackage{amsmath}
\usepackage{graphicx}
\usepackage[dvipsnames,table]{xcolor}
\usepackage{underscore}
\usepackage{float}
\usepackage{tabularx}
\usepackage{booktabs}
\usepackage{minted}
\usepackage[normalem]{ulem}
\usepackage{multirow}
\usepackage{fontawesome}
\usepackage{longtable}
\usepackage{eso-pic}
\usepackage{setspace}
\usepackage{authblk}
\usepackage{geometry}
\usepackage[round,authoryear]{natbib}
\usepackage[colorlinks=true,
            linkcolor=blue,
            citecolor=blue,
            urlcolor=blue]{hyperref}
\usepackage[capitalise]{cleveref}

\newcommand{\APFDc}{\(\text{APFD}_C\)}
\newcommand{\rAPFD}{\(r\text{APFD}\)}
\newcommand{\rAPFDc}{\(r\text{APFD}_C\)}
\newcommand{\cmark}{\textcolor{ForestGreen}{\faCheck}}
\newcommand{\xmark}{\textcolor{Brown}{\faTimes}}

\AddToShipoutPictureFG*{%
  \AtPageUpperLeft{%
    \raisebox{-2cm}{\hspace{3cm}\begin{minipage}{\dimexpr\paperwidth-6cm\relax}{\fontsize{6.5pt}{8pt}\selectfont\copyright 2026. This manuscript version is made available under the CC-BY-NC-ND 4.0 license \url{https://creativecommons.org/licenses/by-nc-nd/4.0/}}\vspace{1mm}\\
    Tomasz Chojnacki and Lech Madeyski, ``Test case prioritization: A systematic review using snowballing and TCPFramework with approach combinators'', \textit{Information and Software Technology}, vol. 194, p. 108062, 2026. DOI: \href{https://doi.org/10.1016/j.infsof.2026.108062}{10.1016/j.infsof.2026.108062}.\end{minipage}}%
  }%
}

\title{Test Case Prioritization: A Systematic Review using Snowballing and TCPFramework with Approach Combinators}

\author[1]{Tomasz Chojnacki%
  \thanks{ORC ID: 0009-0005-0566-2462.
          Email: \texttt{260365@student.pwr.edu.pl}}}
\author[1]{Lech Madeyski\thanks{Corresponding author.
          ORC ID: 0000-0003-3907-3357.
          Email: \texttt{lech.madeyski@pwr.edu.pl}}}
\affil[1]{Wroclaw University of Science and Technology, Wroclaw, Poland}

\date{}

\begin{document}

\maketitle

\begin{abstract}
\noindent \textit{Context:} Test case prioritization (TCP) is a technique
widely used by software development organizations to accelerate regression
testing.\\ \textit{Objectives:} We aim to systematize existing TCP knowledge
and to propose and empirically evaluate a new TCP approach.\\
\textit{Methods:} We conduct a systematic review (SR) using snowballing on
TCP, implement a~comprehensive platform for TCP research (TCPFramework),
analyze existing evaluation metrics and propose two new ones (\rAPFDc{} and
ATR), and develop a~family of ensemble TCP methods called approach
combinators.\\
\textit{Results:} The SR helped identify 324 studies related to TCP. The
techniques proposed in our study were evaluated on the RTPTorrent dataset,
consistently outperforming their base approaches across the majority of
subject programs, and achieving performance comparable to the current state
of the art for heuristical algorithms (in terms of \rAPFDc{}, NTR, and
ATR), while using a distinct approach.\\
\textit{Conclusions:} The proposed methods can be used efficiently for TCP,
reducing the time spent on regression testing by up to 2.7\%. Approach
combinators offer significant potential for improvements in future TCP
research, due to their composability.
\end{abstract}

\noindent\textbf{Keywords:} test case prioritization; regression test
prioritization; continuous integration; ensemble methods; code representation

\section{Introduction}

Software bugs can frustrate users and reduce their satisfaction~\citep{saha2014anempirical} or even cause health and safety risks in certain areas. However, bugs can occur in all software systems~\citep{ajorloo2024asystematic}. To ensure that a program behaves as expected, software testing is performed. In particular, regression testing (RT) is commonly used to ensure that changes to the system did not introduce unwanted behavior~\citep{washizaki2024swebok}.

Due to the volume of executed tests, RT can be costly, both in terms of time and money, with estimates stating that it may comprise between 33\% and 50\% of software expenses~\citep{washizaki2024swebok,myers2011theart,khatibsyarbini2018test}. Regression testing is at the same time an important part of Agile and continuous integration (CI), where automated RT suites are run on each build~\citep{washizaki2024swebok,haghighatkhah2018test,pradolima2020test,marijan2023comparative}. However, CI processes often require rapid turnaround, to avoid delays in build cycles, and to provide software developers with the necessary feedback~\citep{bagherzadeh2022reinforcement,marijan2023comparative}.

Common approaches to speed up the RT process include test suite minimization, test case selection, and test case prioritization~\citep{washizaki2024swebok}, as systematized by \citet{yoo2012regression}. In particular, test case prioritization (TCP), also called regression test prioritization (RTP), aims to reorder test cases to produce the permutation that would detect possible faults as early as possible. It allows all test cases to run in the worst case, but also permits early termination, for example, when the first failing case is encountered~\citep{yoo2012regression}.

As further discussed in \Cref{section:related-work}, over the years, hundreds of TCP methods have been researched. The technique is also used by practitioners, finding success in large software development entities such as Google~\citep{memon2017taming} and Microsoft~\citep{czerwonka2011crane}. However, there are still opportunities for improvement.

Consequently, we have two main objectives. The first is to perform a literature review in the field of TCP. The second and larger aim is to propose and evaluate a new technique for TCP. Our contributions consist of:

\begin{itemize}
    \item A new systematic review using snowballing on the topic of test case prioritization that follows modern review guidelines~\citep{kitchenham2023segress,wohlin2014guidelines} and is one of the largest reviews of the TCP literature in terms of the count of identified works.
    \item Implementation of an open-source, unified TCP approach development and evaluation infrastructure (TCPFramework) that can be used to test various prioritization methods on the state-of-the-art RTPTorrent dataset.
    \item Review and analysis of existing TCP evaluation metrics, including recent proposals, such as \rAPFD{}, RPA, NRPA, and NTR. Proposition of two new TCP evaluation metrics: \rAPFDc{} (to assess the quality of prioritized suites) and ATR (to rate the time effectiveness of approaches).
    \item Introduction of the mental framework of TCP approach combinators, consisting of three method families (mixers, interpolators, and tiebreakers), all of which require no prior training and consistently outperform their base approaches.
    \item Empirical evaluation of proposed prioritizers, which, despite using a distinct method, achieve results comparable to the current heuristic state of the art, and almost three times better than the random approach, reducing the time spent on regression testing by up to a dozen hours.
\end{itemize}

The structure of the paper is presented below. First, after the introduction, the literature review overview and related TCP work are described in \Cref{section:related-work}. Then, a new solution and its implementation are presented during \Cref{section:proposed-approach}. The solution is subsequently evaluated in \Cref{section:evaluation}, with the results listed in \Cref{section:results}. Finally, the approach is discussed alongside further work suggestions in \Cref{section:discussion}, followed by the conclusion.

\section{Related work} \label{section:related-work}

In order to summarize the current state of the art of TCP, and to locate opportunities for further improvement, a systematic review (SR) using snowballing was undertaken. This section describes it in terms of the planning phase, the search procedure, and its results. Then, the most relevant related work is outlined, and an analysis of available TCP evaluation metrics is performed.

\subsection{Literature review process}

Systematic review is a technique to systematically discover, evaluate, and synthesize scientific research related to a specific topic. Compared to traditional unsystematic reviews, systematic reviews are conducted formally, using a set of methodological steps and approaches listed in a review protocol~\citep{kitchenham2007guidelines}. The following review adheres to the SEGRESS guidelines~\citep{kitchenham2023segress} and uses the snowballing method introduced by~\citet{wohlin2014guidelines}.

The systematic review usually begins with the planning phase, where the research questions are listed and the research protocol is written. Expanding on the original motivation, the following research questions were stated:
\begin{itemize}
    \item \textit{(SRRQ1) What are the available datasets and subject programs for TCP?}
    \item \textit{(SRRQ2) What are the state-of-the-art TCP algorithms?}
\end{itemize}

Upon formulating the research objectives, the review protocol was written. According to the available recommendations, the protocol should, among others, describe~\citep{kitchenham2007guidelines}: the motivation for the SR, the previously defined research questions, search methods (including study selection criteria), data extraction, and analysis strategies. The protocol attempts to adhere to the PRISMA-P statement~\citep{moher2015preferred}, as SEGRESS~\citep{kitchenham2023segress} recommends. It should be noted that the protocol was written before the subsequent SR steps were carried out to reduce bias and published online as Section 1 of the online Appendix to this paper\footnote{\url{https://madeyski.e-informatyka.pl/download/ChojnackiMadeyski25Appendix.pdf}\label{fn:Appendix}}.

The following \textbf{inclusion criteria} were stated: 
\begin{enumerate}
    \item the article presents a new TCP dataset (SRRQ1); 
    \item the article describes a new TCP algorithm (SRRQ2);
    \item the article is a secondary study on the topic of TCP;
    \item the article compares available TCP solutions.
\end{enumerate}
Similarly, the following \textbf{exclusion criteria} were stated: 
\begin{enumerate}
    \item the paper places the main emphasis on test case generation, selection, or minimization;\item the paper is unrelated to the Computer Science research area; 
    \item the paper was not peer-reviewed or has not yet been published;
    \item the paper consists of less than~5~pages;\item the paper was not written in the English language.
\end{enumerate}

\begin{figure}[htbp]
    \centering
    \includegraphics[width=0.59\textwidth]{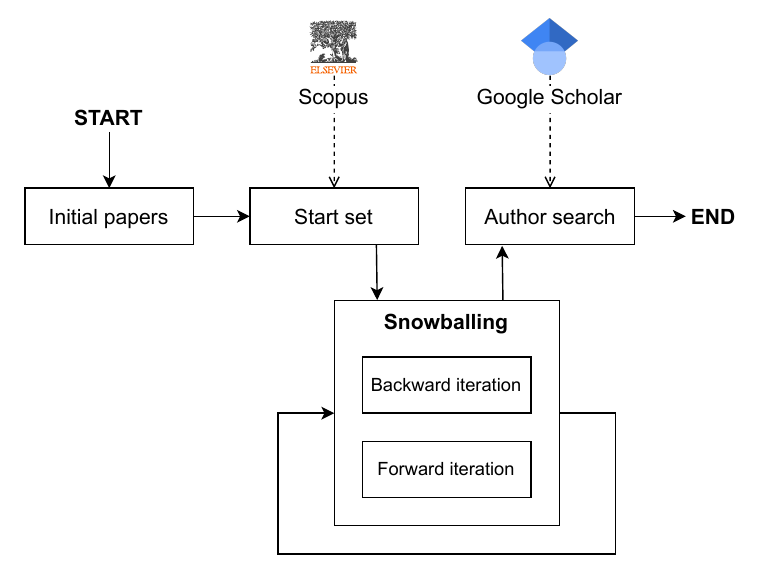}
    \includegraphics[width=0.39\textwidth]{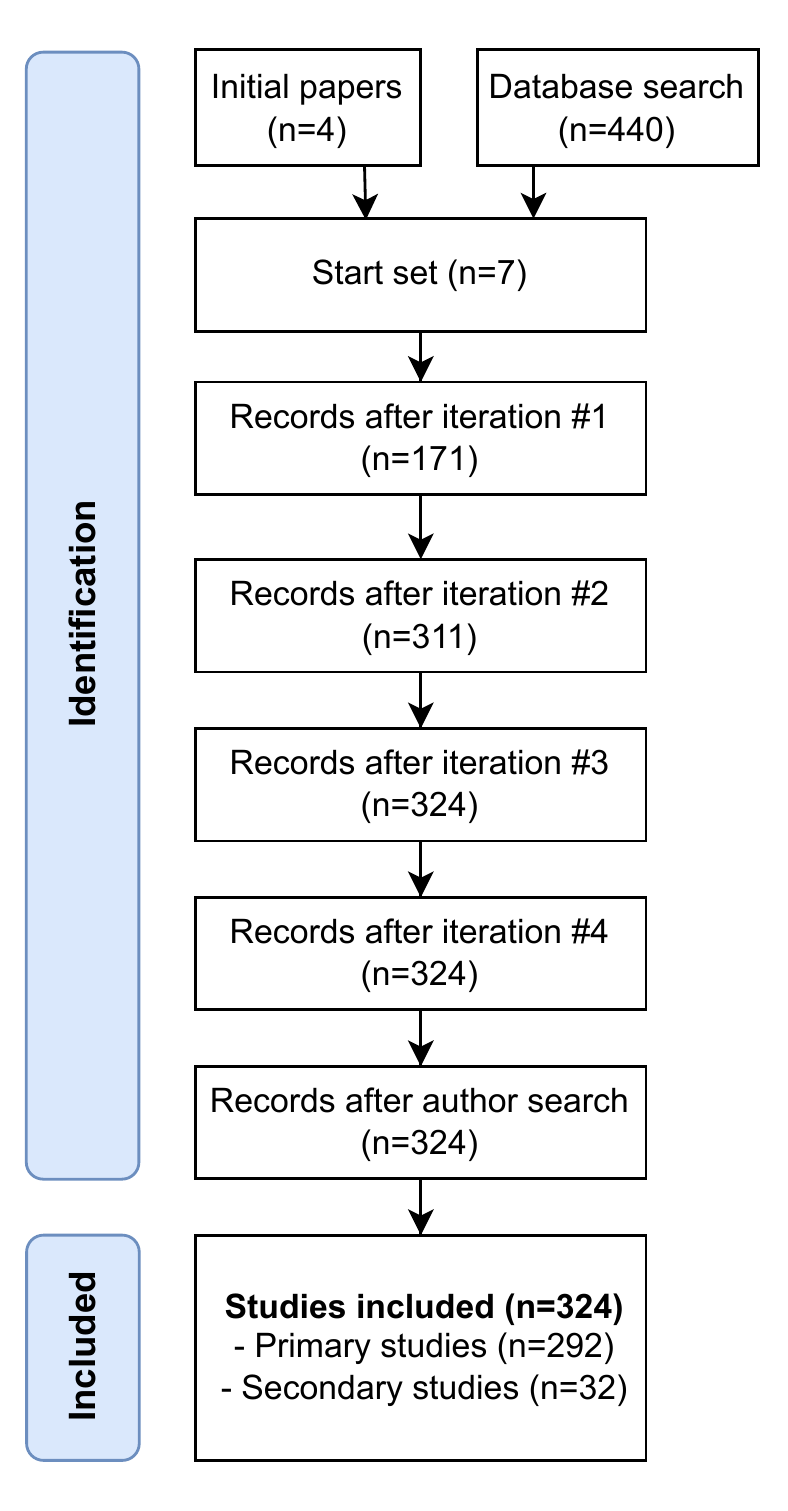}
    \caption{Workflow of the search process and flowchart of paper counts at each stage.}
    \label{fig:snowballing}
\end{figure}

The search process used in the SR follows the snowballing approach recommended by~\citet{wohlin2014guidelines}. Each encountered article is evaluated using the inclusion and exclusion criteria noted in the review protocol linked above. Google Scholar\footnote{\url{https://scholar.google.com}} is used for forward snowballing, the Scopus\footnote{\url{https://www.scopus.com/search/form.uri}} database is used to construct the start set, and the reference lists contained in the articles themselves are used for backward snowballing. \Cref{fig:snowballing} illustrates the implemented search process together with a flowchart summarizing the number of records remaining after each stage.

The presented search process differs from the original snowballing guidelines in two minor respects. These changes were implemented to accommodate the study’s time constraints and do not violate any principles specified in the guidelines.

First, \citet{wohlin2014guidelines} recommends directly contacting all researchers associated with the included papers after the snowballing iterations, while also allowing alternative methods to identify additional studies. As contacting authors was infeasible within the research timeframe, we instead analyzed their publication records. Specifically, we systematically examined the Google Scholar pages of authors appearing at least five times in the paper collection to identify additional relevant articles. This threshold serves as a pragmatic heuristic to focus only on recurring contributors to the field and to reduce noise from one-off publications. We refer to this stage as the \textit{author search}.

Secondly, since we previously gathered a list of unmethodically selected publications closely related to the studied research area, it is used to form the snowballing start set. The list of initial papers contains the works of \citet{yang2023can}, \citet{marijan2023comparative}, \citet{yaraghi2023scalable}, and \citet{bagherzadeh2022reinforcement}. Their keywords were analyzed and used to construct the Scopus search query shown in \Cref{lst:scopus-search}. The \textit{test case prioritization} and \textit{test prioritization} were the most common keywords across initial papers and the \texttt{SUBJAREA}, \texttt{PUBSTAGE} and \texttt{LANGUAGE} filters come directly from the SR inclusion and exclusion criteria. The somewhat arbitrary publication year constraint was introduced to target only recent studies (from the past five years), while also including all initial papers. The search returned a total of 440 documents. Most notably, all initial papers were also returned by the query. This ensures that the filters were not too restrictive.

\begin{listing}[htbp]
\vspace{5mm}\begin{minted}{text}  
( TITLE-ABS-KEY ( "test case prioritization" )
    OR TITLE-ABS-KEY ( "test prioritization" ) )
AND ( LIMIT-TO ( SUBJAREA , "COMP" ) )
AND ( LIMIT-TO ( PUBSTAGE , "final" ) )
AND ( LIMIT-TO ( LANGUAGE , "English" ) )
AND PUBYEAR > 2019
\end{minted}
\caption{The Scopus search string used to find the snowballing start set.}
\label{lst:scopus-search}
\end{listing}

The result list was then manually traversed, using inclusion and exclusion criteria, to identify works that could act as the start set. The guidelines of \citet{wohlin2014guidelines} state that a good start set should include papers from different communities, publishers, years, and authors. In the end, the following articles were selected: \citet{bagherzadeh2022reinforcement}, \citet{huang2020regression}, \citet{vescan2024embracing}, \citet{mukherjee2021survey}. It should be noted that the first entry was previously included in the initial paper set, which is permitted by the SR protocol. Thus, while the start set includes four papers, only three new papers were introduced in this stage. The constructed set spans multiple years, there are no common authors between any two papers in the list, the identified authors come from different communities, and the articles were all published in distinct venues. The selected works contain a total of 274 references and 222 citations (duplicates included), which should provide sufficient resources for the first iteration of snowballing. Subsequently, multiple iterations of snowballing were conducted.

\subsection{Systematic review results}

During the entire snowballing process, a total of 324 articles were identified that comply with the previously specified inclusion and exclusion criteria (and their additional addenda), with 292 primary studies and 32 secondary studies. This was achieved through an analysis of more than 42 thousand references and citations, the vast majority of which were duplicates. The count of papers identified at each stage of the SR is presented in \Cref{tab:sr-stages}, along with the number of entries (references and citations) evaluated.

\begin{table}[htbp]
    \centering
    \caption{The total number of new papers included in each search process stage.}
    \begin{tabular}{l r r}
        \toprule
        \textbf{Review stage} & \textbf{Evaluated entries} & \textbf{Identified papers} \\
        & (duplicates included) & (new and unique) \\ \midrule
        Initial papers & --- & 4 \\
        Start set & 440 & \(^*\) 3 \\
        Iteration \#1: Backward & 274 & 104 \\
        Iteration \#1: Forward & 222 & 60 \\
        Iteration \#2: Backward & 6\,734 & 111 \\
        Iteration \#2: Forward & 21\,393 & 29 \\
        Iteration \#3: Backward & 4\,730 & 5 \\
        Iteration \#3: Forward & 4\,437 & 8 \\
        Iteration \#4: Backward & 425 & 0 \\
        Iteration \#4: Forward & 173 & 0 \\
        Author search & 3\,769 & 0 \\
        \bottomrule
    \end{tabular}
    \label{tab:sr-stages}
    
    \vspace{0.25cm}\footnotesize{\(^*\) -- while there is a total of four start set entries, one is also an initial paper}
\end{table}

\begin{figure}[htbp]
    \centering
    \includegraphics[width=0.66\textwidth]{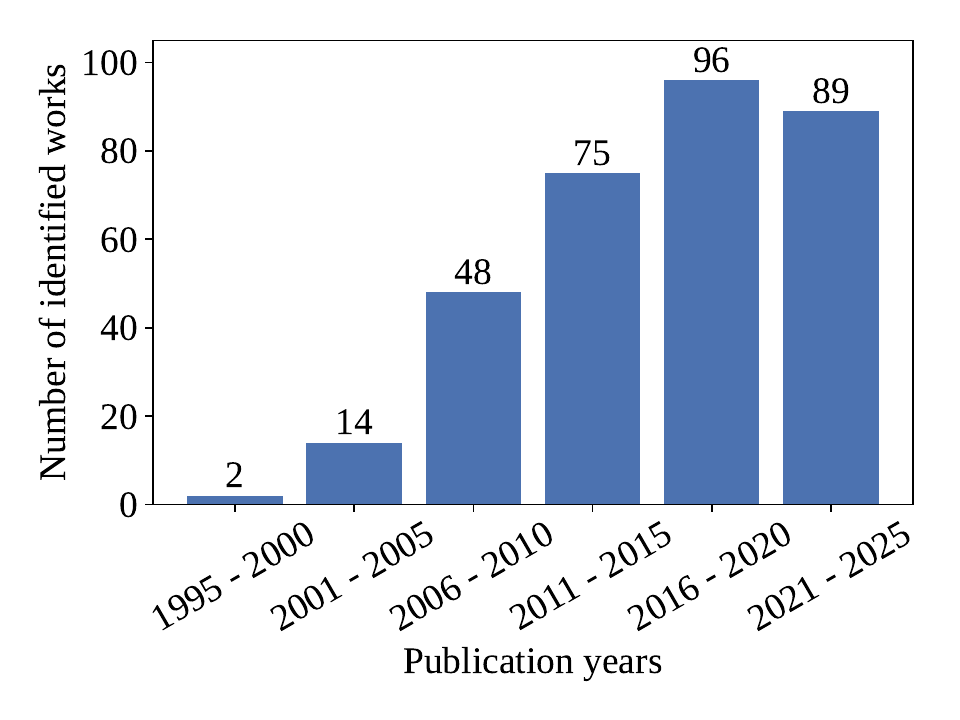}
    \caption{Number of identified works in different publication years.}
    \label{fig:sr-per-year}
\end{figure}

The identified works were spread over multiple decades, with the first published in 1999 by \citet{rothermel1999test} and the last in 2025. The methods used for the SR should ensure that it is unbiased in terms of publication years. The total number of articles from different years is shown in \Cref{fig:sr-per-year}.

A total of 54 articles introduced new datasets, 267 articles presented novel algorithms, 32~articles were secondary studies, and 28 articles were dedicated to empirically comparing existing solutions. Some of the identified surveys cover narrower or wider areas than the proposed SR. Numerous also did not disclose the number of primary studies analyzed. Since the discussed works are not suitable for a~comparison with the proposed SR, of the 32 identified secondary studies, only eight were handpicked and chronologically presented in \Cref{tab:sr-reviews}. Our study reports 292 works rather than 324, as this figure represents only the analyzed primary studies, whereas the total of 324 identified works also includes secondary studies.

\begin{table}[htbp]
    \centering
    \caption{Overview of other surveys encountered during the SR.}
    \resizebox{\textwidth}{!}{
    \begin{tabular}{r l c r}
        \toprule
        \textbf{Year} & \textbf{Study reference} & \textbf{Years covered} & \textbf{Identified works} \\ \midrule
        2012 & \citet{singh2012systematic} & 1997 -- 2011 & 65 \\
        2013 & \citet{catal2013test} & 2001 -- 2011 & 120 \\
        2017 & \citet{decamposjunior2017test} & 1999 -- 2016 & 90 \\
        2018 & \citet{khatibsyarbini2018test} & 1999 -- 2016 & 80 \\
        2020 & \citet{decastrocabrera2020trends} & 2017 -- 2019 & 320 \\
        2020 & \citet{pradolima2020test} & 2009 -- 2019 & 35 \\
        2021 & \citet{mukherjee2021survey} & 2001 -- 2018 & 90 \\
        2023 & \citet{abubakar2023review} & 2002 -- 2021 & 250 \\
        \midrule
        2025 & \textit{proposed} & 1999 -- 2025 & 292 \\
        \bottomrule
    \end{tabular}
    }
    \label{tab:sr-reviews}
\end{table}

In general, the thoroughness of the proposed review is difficult to compare with other studies, as they contained diverse research questions and years covered, and it is possible that some other reviews were missed. Nevertheless, the conducted study represents one of the largest SRs of TCP, in terms of the number of identified publications. Among the analyzed surveys, only \citet{decastrocabrera2020trends} included a higher number of works, when considering raw counts.

The accumulated data was used to answer SRRQ1. Since TCP can be applied to almost any software product where testing is used, nearly any software repository can be adapted to act as a dataset. Some solutions followed this approach, with the most common choice being Software-artifact Infrastructure Repository projects\footnote{\url{https://sir.csc.ncsu.edu}} (SIR): Unix utilities (i.e., bash, grep, sed), Siemens programs, JMeter, JTopas, Ant \citep{pradolima2020test,singh2012systematic,catal2013test,decamposjunior2017test}. Other identified subject programs include compilers (GCC, javac, Jikes), JDepend, Gradebook, Tomcat, Camel, and Drupal.

Some solutions instead opt for processed test suite datasets, such as the Google Shared Dataset of Test Suite Results\footnote{\url{https://code.google.com/archive/p/google-shared-dataset-of-test-suite-results}} (GSDTSR), Defects4j\footnote{\url{https://github.com/rjust/defects4j}}, and Paint Control\footnote{\url{https://bitbucket.org/HelgeS/retecs/src/master/DATA}} \citep{pradolima2020test,decastrocabrera2020trends}. The RTPTorrent dataset is the only identified data source supported by a dedicated research paper~\citep{mattis2020rtptorrent}.

In less complicated problem statements, TCP can be applied to simple \textit{fault matrices}, which state whether a given test case can detect a given fault. An example of such a dataset is the BigFaultMatrix\footnote{\url{https://github.com/dathpo/test-case-prioritisation-ga}}. Finally, synthetically introduced faults (mutants) might be a viable evaluation method for TCP algorithms. \citet{do2006ontheuse} conclude that mutants can be used as a low-cost dataset construction method. However, \citet{luo2018assessing} highlight that TCP algorithms that perform the best on sets of mutants might perform worse on real-world data.

In summary, the answer to SRRQ1 reveals a significant fragmentation in the field: while numerous TCP evaluation datasets exist (SIR programs, GSDTSR, Defects4j, Paint Control, BigFaultMatrix, RTPTorrent, and various individual repositories), no universally adopted benchmark has emerged. This lack of standardization makes cross-study comparisons difficult. Among the identified datasets, RTPTorrent stands out as the only one supported by a dedicated methodological paper that explicitly addresses common dataset quality issues~\citep{mattis2020rtptorrent}. For these reasons, we selected RTPTorrent for our empirical evaluation (see~\Cref{section:evaluation}).

Subsequently, addressing SRRQ2 proved challenging due to the heterogeneity of the field. Our findings align with earlier observations by \citet{singh2012systematic} and \citet{catal2013test}: TCP algorithms are rarely compared under identical conditions (the same datasets, metrics, and definitions), and a significant portion of the proposed methods lack empirical evaluation.

Despite these limitations, some patterns emerge from existing comparative studies. In 2018, \citet{haghighatkhah2018test} analyzed diversity-based and history-based test prioritization applied to six subject programs. \citet{luo2016alarge} instead compared static and dynamic test case prioritization techniques. Most notably, \citet{marijan2023comparative} compared ML approaches with heuristic-based techniques, finding that ROCKET~\citep{marijan2013test} performs comparably to or better than many ML-based methods. The ROCKET technique is also the second most widely used baseline in the evaluation of TCP, after random test case ordering, according to \citet{pradolima2020test}. Based on these findings, we include ROCKET as a primary baseline in our evaluation.

To summarize, the SR revealed that the TCP research landscape is characterized by methodological fragmentation: diverse datasets, incompatible metrics, and limited cross-study comparability. These findings directly informed our research choices: we selected RTPTorrent as the most rigorously documented dataset, adopted ROCKET as a state-of-the-art heuristic baseline, and designed approach combinators as training-free methods that do not depend on specific historical data formats or require extensive tuning.
\Cref{tab:sr-decisions} provides a systematic mapping of how specific SR findings informed key methodological decisions in this work.

\begin{table}[htbp]
    \centering
    \caption{Mapping of SR findings to research decisions.}
    \resizebox{\textwidth}{!}{
    \begin{tabular}{p{3.5cm} p{4.5cm} p{7.25cm}}
        \toprule
        \textbf{Research decision} & \textbf{Related SR publications} & \textbf{How SR informed the decision} \\ \midrule
        Dataset: RTPTorrent & \citep{mattis2020rtptorrent} & SRRQ1 revealed dataset fragmentation (SIR, GSDTSR, Defects4j, Paint Control, BigFaultMatrix, etc.); RTPTorrent is the only dataset with a dedicated methodological paper addressing data quality issues. \\
        \addlinespace
        Baseline: ROCKET & \citep{marijan2013test}, \newline \citep{marijan2023comparative}, \newline \citep{pradolima2020test} & SRRQ2: ROCKET is the second most widely used TCP baseline after random ordering; performs comparably to or better than many ML-based methods. \\
        \addlinespace
        Baseline: DBP & \citep{zhou2021beating} & Identified through SR as a recent heuristic method meeting inclusion criteria for baseline selection. \\
        \addlinespace
        Baseline: DFE & \citep{kim2002ahistorybased}, \newline \citep{mattis2020rtptorrent} & Used as a non-trivial baseline in RTPTorrent; represents simple exponential-smoothing TCP approaches. \\
        \addlinespace
        Approach \newline combinators (RQ1) & \citep{huang2021alearntorank}, \newline \citep{cao2022ensemble}, \newline \citep{mondal2021hansie}, \newline \citep{hassan2024anexploratory} & SR identified a gap: prior attempts at combining TCP methods were limited to ML-based approaches (requiring training) or simple voting schemes; no training-free combinators existed. \\
        \addlinespace
        Metric: NTR & \citep{pradolima2022amultiarmed}, \newline \citep{zhao2023revisiting} & SR identified NTR as suitable for evaluating CI applicability; inspired the proposed ATR metric addressing its limitations. \\
        \addlinespace
        Metric: \rAPFD{} & \citep{zhao2023revisiting} & SR identified limitations of APFD (inconsistent value ranges); \rAPFD{} addresses this; inspired the proposed \rAPFDc{} metric. \\
        \bottomrule
    \end{tabular}
    }
    \label{tab:sr-decisions}
\end{table}

In the following sections, we present the key theoretical topics relevant to this work, drawing primarily on information obtained from the literature review.

\subsection{Foundational works on TCP}

To the best of our knowledge, the first mention of TCP occurs in the work of \citet{wong1997astudy}, where it is used as an additional step after test case selection. The first comprehensive look at TCP was then provided by \citet{rothermel1999test}, where a wide range of prioritization methods were investigated.

The test case prioritization problem was previously formally defined in the work of \citet{rothermel2001prioritizing}, and the definition can be seen in \Cref{eq:tcp-problem}:
\begin{equation} \label{eq:tcp-problem}
\begin{aligned}
    \text{\textbf{Given:} } & T\text{, a test suite; }PT\text{, the set of permutations of }T\text{;} \\
    & f\text{, a function from }PT\text{ to the real numbers.} \\
    \text{\textbf{Problem:} } & \text{Find }T' \in PT\text{ such that }\left(\forall\,T''\right)\left(T'' \in PT\right) \\
    & \left(T'' \neq T'\right) \left[f\left(T'\right) \geq f\left(T''\right)\right]\text{.}
\end{aligned}
\end{equation}

In the problem statement, \(f\) is used to evaluate the quality of the ordering (higher values are better). In simpler words, our aim is to find the best permutation of test cases in terms of a given metric.

\begin{figure}[htbp]
    \centering
    \includegraphics[width=\textwidth]{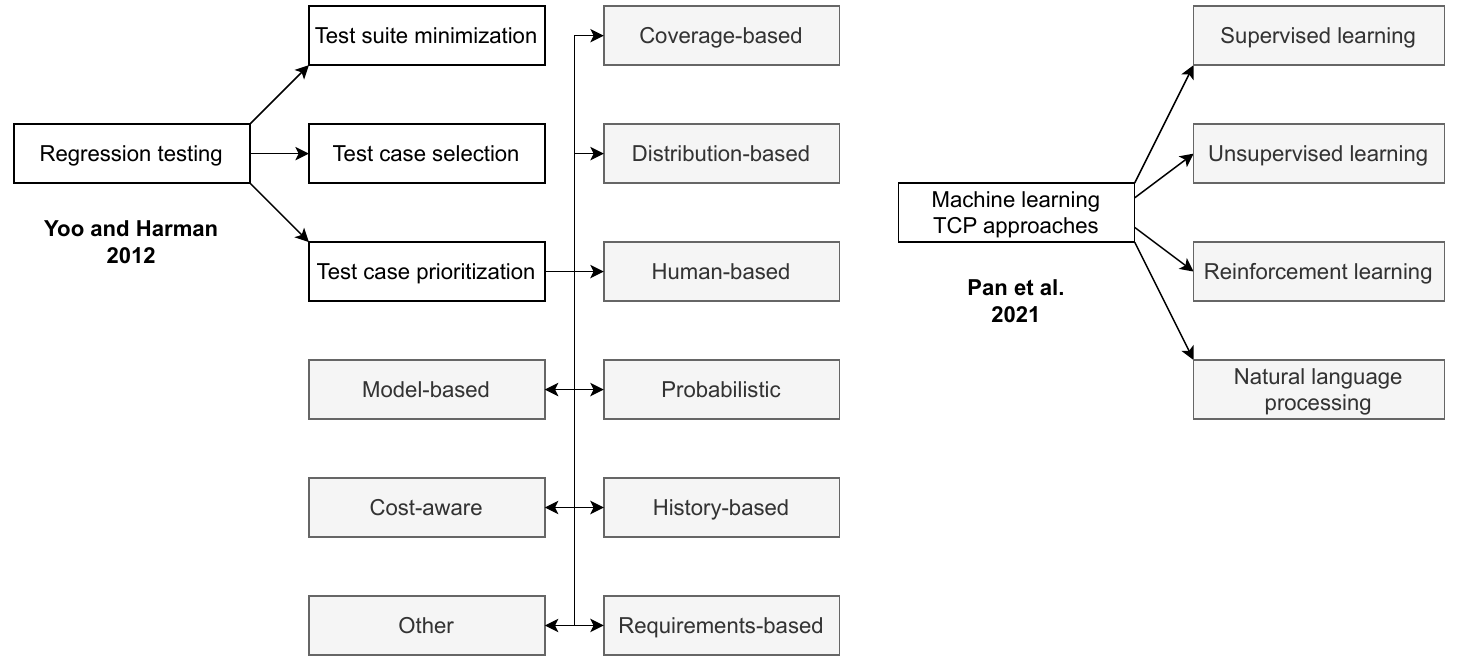}
    \caption{Taxonomies of TCP (on a basis of~\citep{yoo2012regression}) and ML-based TCP (on a basis of~\citep{pan2021test}) approaches.}
    \label{fig:tcp-taxonomy}
\end{figure}

Over the years, multiple TCP approach categorizations have been proposed. Examples include the taxonomies used by \citet{khatibsyarbini2018test}, \citet{singh2012systematic}, and \citet{mukherjee2021survey}. For the purpose of the study, the classification introduced by \citet{yoo2012regression} was selected. It includes the following categories of TCP approaches~\citep{yoo2012regression}: coverage-based, distribution-based, human-based, probabilistic, history-based, requirements-based, model-based, cost-aware, and other approaches. Using insights collected during the SR, it can be observed that machine learning solutions are underrepresented in the list of approaches. \citet{pan2021test} provided a separate taxonomy for just ML-based approaches, with the following categories: supervised learning, unsupervised learning, reinforcement learning, and natural language processing. Both described classifications are presented side by side in \Cref{fig:tcp-taxonomy}.

In addition to the taxonomies in \Cref{fig:tcp-taxonomy}, the TCP approaches are characterized by their properties. Most existing TCP methods can be classified as either static or dynamic~\citep{luo2016alarge}. Static techniques operate only on the source and test code, using only static program analysis. However, dynamic techniques utilize rich run-time semantics. This enables the algorithm to dynamically reorder the cases as the tests are executed. The techniques can also be divided into black-box and white-box types~\citep{henard2016comparing}. Black-box approaches do not require access to the source code of the system under test; in contrast, white-box approaches can use the source code information to improve the prioritization performance.

\subsection{Approach combinators}

After an analysis of the titles and full abstracts of all publications identified during the undertaken systematic review, only two works adjacent to the methods proposed in the work were identified.

\citet{huang2021alearntorank} attempted to combine the results of multiple heuristic TCP methods using ML algorithms. They focused on model-based prioritization and collected six simple estimators rating test cases based on state transitions. The outputs from particular heuristics are then accumulated using a random forest model to produce the final prediction. Their research differs from the approach we propose in this paper in the following ways: it focuses only on model-based TCP, it works only on top of certain methods, and it requires prior training.

\citet{hassan2024anexploratory} investigated the use of random sorting to resolve the ties in the ranking of the test cases. They concluded that this method does not yield optimal results and should only be used as a final resort. The approach proposed in this paper supports random tie-breaking, but it also offers more sophisticated strategies.

An additional ad hoc search was performed, after the solutions from this work had already been implemented, in an attempt to find similar works in the literature. Two additional articles, described below, were identified.

The research of \citet{cao2022ensemble} used a simple voting scheme to combine the results of multiple models. Each model assigned votes in a decreasing sequence, and the final score was calculated as the sum of the votes and used to determine the final result. Their research differs from the proposed model in the following ways: they focused only on user interface testing, the combination used the simplest voting scheme available, and no method to assign different weights to models was presented.

The work of \citet{mondal2021hansie}, who introduced the Hansie framework for TCP, is most adjacent to our proposed solutions. They assumed that internal prioritizer mechanisms are unknown. Then, the Borda count is used for result combination, similarly to one of the proposed approach combinators. Despite resemblances, we believe that their technique is not equivalent. The Borda count algorithm constitutes only a minor part of the research conducted, and Hansie uses parallel prioritization to resolve ties, while the proposal introduces different methods to deal with ties.

\subsection{Metrics for TCP evaluation}

\citet{rothermel1999test} introduced the \textit{average of the percentage of faults detected} (APFD) metric in 1999~\citep{lou2019asurvey,zhao2023revisiting}. Although it exhibits multiple limitations, it is still the most widely used TCP metric in the literature~\citep{abubakar2023review,decamposjunior2017test,catal2013test,singh2012systematic,khatibsyarbini2018test,mukherjee2021survey}. APFD measures how quickly an ordered test case list detects faults. The metric ranges from 0~to~1, with higher values of the metric representing better orderings~\citep{elbaum2000prioritizing,catal2013test,mukherjee2021survey,zhao2023revisiting,lou2019asurvey}. It can be calculated using the formula from \Cref{eq:metrics-apfd}~\citep{lou2019asurvey,wang2020arevisit,zhao2023revisiting}:
\begin{equation} \label{eq:metrics-apfd}
\text{APFD}(s) = 1 - \frac{\sum_{j=1}^m \text{TF}_j}{n m} + \frac{1}{2 n},
\end{equation}
\noindent where \(s\) is the suite, \(n\) is the number of its test cases, \(m\) is the number of identified faults, and \(\text{TF}_j\) is the rank of the first test in the ordered suite, which reveals the \(j\)-th fault.

The \textit{cost-cognizant weighted average percentage of faults
detected} (\APFDc{}) introduced by \citet{elbaum2001incorporating} extends APFD to incorporate differences in fault severities and test case costs (for example, their execution times). The \APFDc{} metric is very popular and is sometimes considered the second most widely used TCP metric~\citep{decamposjunior2017test,singh2012systematic}. It can be seen as an extension of APFD, where test case costs and fault severities are known, and can be calculated as shown in \Cref{eq:metrics-apfdc}~\citep{wang2020arevisit,lou2019asurvey}:
\begin{equation} \label{eq:metrics-apfdc}
    \text{APFD}_C\left(s\right) = \frac{\sum_{j=1}^m \left(f_j \cdot \left(\sum_{i=\text{TF}_j}^n t_i - \frac{1}{2} t_{\text{TF}j}\right)\right)}{\sum_{j=1}^n t_j \cdot \sum_{j=1}^m f_j},
\end{equation}
\noindent where \(s\), \(n\), \(m\), \(\text{TF}_j\) have the same meaning as in \Cref{eq:metrics-apfd} and \(f_i\) and \(t_j\) describe the severity of \(i\)-th fault and execution cost of \(j\)-th test case, respectively. It has the same value range as APFD~\citep{wang2020arevisit}. In real-world studies, fault severities are rarely known. The \APFDc{} formula also works when all severities are assumed to be equal~\citep{elbaum2001incorporating,wang2020arevisit,lou2019asurvey}.

Whenever we introduce constraints to TCP, some faults might not get detected, as some test cases are not run. \citet{qu2007combinatorial} introduced \textit{normalized APFD} (NAPFD) in 2007, including an additional \(p\) factor that represents the fraction of identified faults in all identifiable faults~\citep{abubakar2023review,lou2019asurvey,wang2020arevisit,zhao2023revisiting}. The formula for NAPFD is given in \Cref{eq:metrics-napfd}, where \(\text{TF}_j\) is set to zero if a fault is never detected~\citep{lou2019asurvey,wang2020arevisit}:
\begin{equation} \label{eq:metrics-napfd}
    \text{NAPFD}\left(s\right) = p - \frac{\sum_{j=1}^m \text{TF}_j}{nm} + \frac{p}{2n}.
\end{equation}

\citet{zhao2023revisiting} identified weaknesses in existing metrics from the APFD family, namely that APFD has different value ranges depending on fault and test case counts. This also means that although it returns results bounded by 0 and 1, the optimal solution does not have an APFD of 1, and neither does the worst solution have an APFD of 0. They addressed this weakness, min-max mapping the measure, introducing the \textit{rectified APFD} (\rAPFD{}), which ensures that the optimal solution has \(r\text{APFD} = 1\). The value of \rAPFD{} for a given test suite is shown in \Cref{eq:metrics-rapfd}:
\begin{equation} \label{eq:metrics-rapfd}
    r\text{APFD}\left(s\right) = \frac{\text{APFD}\left(s\right) - \text{APFD}_{\min}\left(s\right)}{\text{APFD}_{\max}\left(s\right) - \text{APFD}_{\min}\left(s\right)}.
\end{equation}

\citet{bertolino2020learningtorank} introduced the \textit{rank percentile average} (RPA) and its normalization called \textit{normalized-rank-percentile-average} (NRPA) to evaluate the order of the test suite independently of the algorithm and the choice of criteria~\citep{zhao2023revisiting,bagherzadeh2022reinforcement}. These measures have been criticized before~\citep{zhao2023revisiting,bagherzadeh2022reinforcement}. \citet{bagherzadeh2022reinforcement} note that NRPA treats successful and failed test cases equally, while failures are generally more important. \citet{zhao2023revisiting} support the limitation described previously and additionally report that sometimes worse sequences report higher NRPA values.

The applicability of TCP is also evaluated in the literature. If a method has a high cost, it might be really effective, but not applicable in practical contexts. Various time-related metrics are most commonly used for this goal~\citep{khatibsyarbini2018test,pradolima2020test}. \citet{pradolima2022amultiarmed} defined the \textit{normalized time reduction} (NTR) metric to measure time savings gained if the cycle stopped at the first detected fault. It can be determined using \Cref{eq:metrics-ntr}~\citep{pradolima2022amultiarmed,zhao2023revisiting}:
\begin{equation} \label{eq:metrics-ntr}
    \text{NTR} = \frac{\sum_{i=1}^{\text{CI}^{\text{fail}}} \left(\hat{r}_i - r_i\right)}{\sum_{i=1}^{\text{CI}^{\text{fail}}} \hat{r}_i},
\end{equation}
\noindent where \(\text{CI}^\text{fail}\) is the number of failed cycles, \(\hat{r}_i\) is the execution time of \(i\)-th failing cycle, and \(r_i\) is the time to first fault of \(i\)-th failing cycle. The NTR measure was later adopted by \citet{zhao2023revisiting} to evaluate the applicability of TCP models.

All previously described measures are compared in \Cref{tab:effectiveness-measure-comparison}.

\begin{table}[htbp]
    \centering\footnotesize
    \caption{Comparison of described TCP effectiveness metrics.}
    \resizebox{\textwidth}{!}{
    \begin{tabular}{l c c c c c c}
    \toprule
    \multirow{2}[3]{*}{\textbf{Measure}} & \multirow{2}[3]{*}{\textbf{Computable}} & \multirow{2}[3]{*}{\textbf{Normalized} \(^*\)} & \multicolumn{4}{c}{\textbf{Awareness}} \\ \cmidrule{4-7}
 &  &  & \textbf{Failure} & \textbf{Cost} & \textbf{Severity} & \textbf{Constraint} \\ \midrule
    \textbf{APFD}~\citep{rothermel1999test} & \textcolor{Brown}{failed} cycles & \xmark{} & \cmark{} & \xmark{} & \xmark{} & \xmark{} \\
    \textbf{\APFDc{}}~\citep{elbaum2001incorporating} & \textcolor{Brown}{failed} cycles & \xmark{} & \cmark{} & \cmark{} & \cmark{} & \xmark{} \\
    \textbf{\APFDc{}} {\footnotesize (simple)} & \textcolor{Brown}{failed} cycles & \xmark{} & \cmark{} & \cmark{} & \xmark{} & \xmark{} \\
    \textbf{NAPFD}~\citep{qu2007combinatorial} & \textcolor{Brown}{failed} cycles & \xmark{} & \cmark{} & \xmark{} & \xmark{} & \cmark{} \\
    \textbf{\rAPFD{}}~\citep{zhao2023revisiting} & \textcolor{Brown}{failed} cycles & \cmark{} & \cmark{} & \xmark{} & \xmark{} & \xmark{} \\
    \textbf{RPA}~\citep{bertolino2020learningtorank} & \textcolor{ForestGreen}{all} cycles & \xmark{} & \xmark{} & \xmark{} & \xmark{} & \xmark{} \\
    \textbf{NRPA}~\citep{bertolino2020learningtorank} & \textcolor{ForestGreen}{all} cycles & \cmark{} & \xmark{} & \xmark{} & \xmark{} & \xmark{} \\
    \textbf{NTR}~\citep{pradolima2022amultiarmed} & entire history & \cmark{} & \cmark{} & \cmark{} & \xmark{} & \xmark{} \\ \bottomrule
    \end{tabular}
    }
    \label{tab:effectiveness-measure-comparison}
        
    \vspace{0.25cm}\footnotesize{\(^*\) -- whether for a metric \(M\), there occurs \(M_{\max} = 1\)}
\end{table}

\section{Methods} \label{section:proposed-approach}

The proposed solution consists of both a mental framework and a concrete Python implementation of the described methods. This section starts with a~description of the research framework. Then, different non-combinator models are outlined. Finally, the approach, combinator strategy, is introduced, and concrete examples of models built using combinators are presented.

\subsection{Research infrastructure (TCPFramework)}

Although many works in the field of TCP publish associated software packages to aid reproducible research \citep{mondal2021hansie,zhao2023revisiting,bertolino2020learningtorank,luo2016alarge,henard2016comparing}, most of them are ad hoc scripts used to collect results for research questions. We were unable to find a comprehensive and unified framework for the development and evaluation of TCP approaches in the existing literature. To aid further research, both for the scope of this research and for other researchers, we propose an extensible platform for the development of test case prioritization methods, called TCPFramework.

The system is built as a Python module and allows for extension through subclassing and interface implementation. The \texttt{Approach} abstract base class is used as a template for all TCP methods. The type hierarchy ensures safety guarantees, such as: no test case can be marked as prioritized twice; all input test cases must be present in the output; test cases not present in the cycle cannot be prioritized; the case information (execution time, verdict) is not extractable before execution. For example, the optimal solution is, correctly, not implementable as a subclass of \texttt{Approach}, since it requires a posteriori cycle information to work. Approaches that dynamically adjust the order based on the execution results of previous test cases are implementable, and a convenient way to express ties in the results is available.

The \texttt{Dataset} class is another form of abstraction offered by the TCPFramework. Its default implementation extracts the necessary information from the RTPTorrent dataset, which is described in further detail in \Cref{section:evaluation}. Other subject programs with compatible data formats can also be loaded, and the \texttt{Dataset} class can be further derived to adapt to other data sources. Furthermore, TCPFramework provides the tools necessary for the calculation of evaluation metrics, in the form of the \texttt{MetricCalc} class. It contains methods that calculate all the metrics covered in this work.

To confirm the validity of TCPFramework and to provide simple baselines for further evaluations, basic approaches based on foundational publications in TCP and straightforward intuitions were developed. These include:
\begin{itemize}
    \item \textit{BaseOrder}: The base order approach executes the test cases from the suite in their original arrangement~\citep{rothermel1999test}. This method can be used as a "no prioritization performed" baseline to evaluate the gains of applying TCP to a certain software project.
    \item \textit{RandomOrder}: The random approach unmethodically shuffles the test suite and executes test cases in the resulting order~\citep{rothermel1999test}.
    \item \textit{RecentnessOrder}: The recentness order rates test cases on the basis of how recently they were introduced to the project. A counter is kept for each case, accumulating the times it was seen in a cycle. The test cases that appeared fewer times in previous regression testing cycles are executed first. This reflects the intuition that verdicts of new test cases are less predictable (since less historical information can be used to estimate their outputs).
\end{itemize}

Other methods are history-based and rate test cases using aggregations of their previous performances. \textit{Exponential smoothing} was also previously used to aggregate historical results for use in TCP~\citep{kim2002ahistorybased,mattis2020rtptorrent}. It is defined using the formula from \Cref{eq:exponential-smoothing}:
\begin{equation} \label{eq:exponential-smoothing}
    P(k) = \begin{cases}
        0, \quad & k = 0 \\ \alpha \cdot f(k) + \left(1 - \alpha\right) \cdot P(k-1), \quad & k \geq 1
    \end{cases}
\end{equation}
\noindent where \(f(k)\) is equal to the value of some property of the execution result for the test case in cycle \(k\) and \(\alpha\) is a~constant used to indicate the importance of recent results. Techniques from this family are outlined below:
\begin{itemize}
    \item \textit{FoldFailsOrder}: The failure folding approach applies an aggregation function to failure counts of particular test cases from each subsequent cycle. The most notable folders include summation and exponential smoothing. Demonstrated fault effectiveness (DFE), previously used by \citet{kim2002ahistorybased} is a special case of this approach.
    \item \textit{ExeTimeOrder}: The execution time order applies exponential smoothing to subsequent execution times of a given test case achieved in prior cycles.
    \item \textit{FailDensityOrder}: The failure density approach combines two previous techniques by applying exponential smoothing separately to both failures and execution time to then prioritize according to their quotient.
\end{itemize}

\subsection{Code representation approach (CodeDistOrder)}

The next developed approach, \textit{CodeDistOrder}, builds on the work of \citet{yang2023can} and \citet{zhou2024ccore}. Code representation aims to embed the rich information present in the source of a program as a~numeric vector~\citep{yang2023can,hrkut2023current}. While the work of \citet{yang2023can} addressed information retrieval-based test prioritization and \citet{zhou2024ccore} used code representation to cluster test cases, the approach presented in this paper instead focuses on the distances between the code representation vectors. These distance measures include the Manhattan distance, the Euclidean distance, and cosine similarity~\citep{hrkut2023current}.

The goal of the proposed optimizer is to find the longest path, in terms of the source code representation distance, going through all test cases. Finding an exact solution to such a~problem is computationally difficult~\citep{ledru2012prioritizing}. Instead, we employ a greedy algorithm that selects the highest possible distance at each step.

\subsection{Approach combinators}

Our main contribution is the development of approach combinators for TCP. The existing literature often employs hybrid approaches that merge multiple information factors. Despite this, the methods found during the SR are characterized by tight coupling between different factors. We observe a need for a universal approach to the merging and enriching of existing TCP techniques. To facilitate the combination of different TCP approaches, regardless of their complexity and implementation details, we believe that they should be treated as black boxes that prioritize cases using opaque mechanisms.

This guideline is respected by all the techniques introduced below, collectively called \textit{approach combinators}. They are a class of TCP approaches that take other approaches, use them to prioritize test suites, and then merge and enrich them. Three kinds of approach combinators are presented: \textit{mixers}, which take multiple prioritization algorithms and mix their weighted results into the final sequence; \textit{interpolators}, which shift the relative importance of different techniques as cycles pass; and \textit{tiebreakers}, that aim to resolve ties produced by an approach, using a different approach or some heuristic.

All developed strategies require no prior training, in contrast to many of the recently proposed ML-based approaches. This makes them applicable even in smaller projects, where there is insufficient historical information. Additionally, smaller development entities can use the methods as they are suitable for low-specification hardware.

\subsection{Mixers}

\begin{figure}[htbp]
    \centering
    \includegraphics[width=\textwidth]{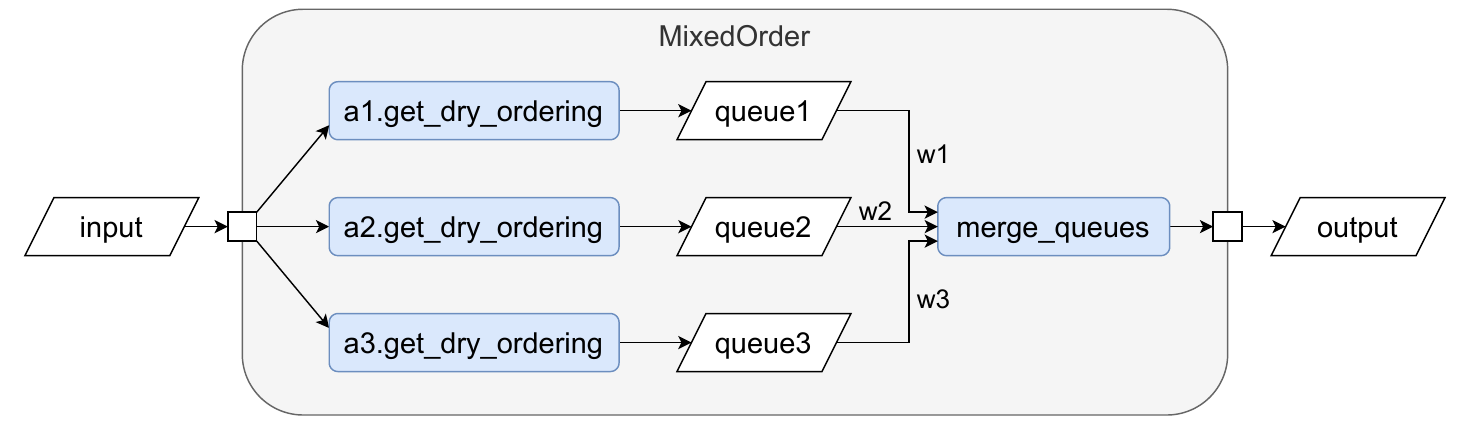}
    \caption{Summary of the general workflow of a mixer.}
    \label{fig:combinators-mixer}
\end{figure}

Mixers take multiple prioritization algorithms and mix their weighted results, returning the final test case sequence. The primary intuition is that various test case properties should be incorporated into an algorithm to produce high-quality predictions. New approaches are produced, which for each cycle prioritize the test suite independently using each sub-approach and then merge the results, called \textit{queues}, into one final list using some algorithm. No internal information leaks from any sub-algorithm. We illustrate our process in \Cref{fig:combinators-mixer}.

The only remaining detail is the way the results are merged. The simplest proposed approach, further called \textit{RandomMixedOrder} takes the first test case from the \(i\)-th queue, where \(i\) was chosen randomly according to the specified weights. Other proposed merging schemes come directly from electoral system theory. Firstly, \textit{BordaMixedOrder} uses the Borda method for positional voting to arrange the suite~\citep{brandt2016handbook}. For each sub-approach outcome, we assign points to test cases based on their position. The first case in a given queue gets \(n - 1\) points (where \(n\) is the size of the suite), the second gets \(n - 2\), etc. until the penultimate and last test cases get 1~and~0~points, respectively. If two test cases are tied, their points are averaged. The points are then multiplied by the weight of a~given sub-method and summed up over all queues for each case. The final order is determined by the sum of the scores of all the cases. The Schulze method is also implemented, as \textit{SchulzeMixedOrder}. It is much more robust than the Borda voting system, providing certain desirable qualities, such as the \textit{majority winner} rule -- if a test case has more than 50\% points, it must win \citep{brandt2016handbook}. It should be noted that, as a variant of the Floyd-Warshall algorithm is used, the computational complexity of this combinator reaches \(O\left(n^3\right)\), making it unsuitable for larger collections of test cases.

We selected these three aggregation strategies to represent a spectrum of complexity and theoretical guarantees. The random baseline serves as a control condition, enabling assessment of whether more sophisticated methods provide meaningful improvements. The Borda count was chosen for its simplicity and low computational overhead, and because it has been successfully applied to TCP by \citet{mondal2021hansie} in the Hansie framework. The Schulze method~\citep{Schulze11}, a Condorcet-consistent approach, was included to evaluate whether stronger theoretical properties (such as the majority winner criterion) translate to better TCP performance, despite its higher computational cost (\(O(n^3)\)). This selection allows us to investigate the trade-off between computational efficiency and aggregation quality in the context of test case prioritization.

In TCPFramework, all mixers implement the \texttt{MixedOrder} abstract base class. As an optimization, if a certain sub-approach has the weight of zero, its order is not computed, as all merging schemes should ignore queues produced by these approaches. Prioritization feedback and the reset action are further propagated to all target methods.

For example, consider a RandomMixedOrder mixer combining ExeTimeOrder with a weight of 0.75 and RecentnessOrder with a weight of 0.25. When a test suite is provided, the mixer first sorts the test cases according to the rules of its sub-approaches. As a result, two permutations of the same test suite are produced: one ordered by prior execution times, and the other ordered by recentness. To produce the final order, the mixer repeatedly selects and runs the first unexecuted test case from one of the two lists at random, with the list based on execution-time having three times the probability of being chosen. This process continues until either all test cases in the suite have been executed or a failing case is encountered.

\subsection{Interpolators}

Interpolators work with different approaches and shift their relative importance values as cycles (failed or any) pass. The main motivation in this case is that numerous simple approaches are history-based and thus perform poorly in the early iterations. However, once they collect enough historical information, they can become fairly performant.

The approach works as follows: two approaches (called \textit{before} and \textit{after}) and a certain cycle count goal (called \textit{cutoff}) are manually set, and the two approaches are mixed according to the current progress towards this objective. So in the first cycle, only the \textit{before} method is executed, and in every cycle after the cutoff, only the \textit{after} technique is run. It should be noted that both approaches can react to execution feedback, regardless of which one was used, meaning that the \textit{after} prioritizer can train itself in earlier cycles. This strategy, called \textit{InterpolatedOrder}, can be conveniently implemented in terms of mixers.

For example, consider an interpolator configured with RandomOrder as the \textit{before} method, FailDensityOrder as the \textit{after} method, and a cutoff value of two. During the first cycle, the entire test suite is prioritized solely using the random prioritizer. In the second cycle, the resulting ordering is an equal combination of the random ordering and the ordering produced by FailDensityOrder. From the third cycle onward, test cases are prioritized exclusively based on their fail density. Despite not being applied in the first cycle and only partially in the second, the FailDensityOrder prioritizer has access to the complete results of both cycles.

\subsection{Tiebreakers}

Tiebreakers resolve ties from an input ordering using another prioritizer or some heuristic. Their construction is motivated by the fact that some approaches naturally produce clustered results. We propose two tiebreakers. Firstly, the \textit{GenericBrokenOrder} is a universal tiebreaker that runs a first sub-approach producing a clustered test case list, and then executes a~second sub-approach to prioritize each group, producing the final order. If the second sub-method returns ties, they are also present in the final output. Similarly to mixers and interpolators, the execution feedback and reset signal are propagated to both sub-approaches.

The second proposed tiebreaker, \textit{CodeDistBrokenOrder}, uses code representation distances to break ties within groups of equal priority. Instead of simply running the CodeDistOrder prioritizer within each cluster, we keep shared sets of prioritized and remaining test cases, ensuring that the chosen candidates are placed as far as possible within the code representation space of the entire test suite.

Tiebreakers can also perform a role similar to that of interpolators. In early cycles, some techniques return a single cluster that contains all test cases. In such a case, when using a tiebreaker, the entire prioritization is delegated to the second sub-approach.

For example, consider a GenericBrokenOrder configured with RecentnessOrder as the primary prioritizer and RandomOrder as the secondary one. Newly introduced test cases in a given cycle are guaranteed to be executed first. When multiple new test cases are added simultaneously, they receive identical scores according to the recentness criterion. Because historical information is not available for these cases, ties are resolved using random ordering. Without this secondary prioritizer, using RecentnessOrder alone would result in the new test cases being executed in their original order.

\subsection{Example combinator models}

Since all techniques described in the previous section are actually entire families of approaches, some concrete, usable examples are given below. To construct a~high-performing prioritizer, we intuitively want to combine multiple information factors. Subsequently, the following sample combined approaches are defined:
\begin{itemize}
    \item \textit{P1} -- a mixer that blends FoldFailsOrder, RecentnessOrder, and ExeTimeOrder. Available in all mixer variants, that is, random (P1.1), Borda (P1.2), and Schulze (P1.3). As we believe execution time to be less impactful, its priority is set to be twice as low.
    \item \textit{P2} -- an interpolator that uses the FailDensityOrder after five failed cycles and an equal mix of ExeTimeOrder and RecentnessOrder before. This also demonstrates the ability to nest approach combinators within each other, as a Borda mixer is used inside an interpolator.
    \item \textit{P3} -- a tiebreaker that resolves ties in the total-strategy FoldFailsOrder prioritizer, using ExeTimeOrder (P3.1) or CodeDistOrder (P3.2).
\end{itemize}

For P1, we predict that combining the failure and cost information factors, similar to FailDensityOrder, will result in better performance than using either factor individually. We additionally use the latest-only RecentnessOrder approach to increase the priority of new test cases. For P2, we observe that FailDensityOrder performs well but is extremely unstable in the first iterations. To perform early prioritization, ExeTimeOrder could be used; however, it also needs a few cycles to stabilize its internal state. We therefore use its mix with a~generally stable RecentnessOrder until a couple of failing cycles are encountered. Finally, for P3, the total strategy of FoldFailsOrder naturally produces rather large clusters, especially in the early cycles. We predict that breaking these tied groups would significantly boost the effectiveness of the approach.

The proposed sample models use several existing TCP techniques. As we want the models to be lightweight, we almost exclusively use methods based on exponential smoothing, first proposed by \citet{kim2002ahistorybased}. We list all choices and justify them in \Cref{tab:example-model-refs}. Further implementation details for both the combinators and the concrete models can be found in the online appendix\hyperref[fn:Appendix]{\textsuperscript{\ref*{fn:Appendix}}} and the reproduction package\footnote{\url{https://github.com/LechMadeyski/MSc25TomaszChojnacki}\label{fn:Reproduction}}.

\begin{table}[htbp]
    \centering
    \caption{Existing TCP methods used within the proposed sample models.}
    \begin{tabular}{l p{3.25cm} p{6.75cm}}
        \toprule
        \textbf{Model} & \textbf{Existing methods} & \textbf{Justification} \\ \midrule
        \textbf{P1} & FoldFailsOrder \par ExeTimeOrder \par RecentnessOrder & We want to prioritize cases based on three criteria: historical failures, execution efficiency, and expected result variability. We thus promote cases that failed in previous executions (FoldFailsOrder), are quick to execute (ExeTimeOrder), and are likely to produce unstable results due to their recent introduction (RecentnessOrder). \\ \midrule
        \textbf{P2} & FailDensityOrder \par ExeTimeOrder \par RecentnessOrder & FailDensityOrder tends to perform well in later testing cycles; however, it is unstable in early cycles, as it relies on the ratio of two estimated averages. We thus want to interpolate it with prioritizers that demonstrate stable performance early on. In our preliminary experiments, ExeTimeOrder and RecentnessOrder both have this quality. \\ \midrule
        \textbf{P3} & FoldFailsOrder \par ExeTimeOrder \par CodeDistOrder & The total strategy of FoldFailsOrder yields large clusters of cases with identical scores, which limits its effectiveness. We hypothesize that breaking these ties is beneficial. Prioritizers that produce distinct scores are best for tiebreaking. Both ExeTimeOrder and CodeDistOrder satisfy this property and are thus well-suited. \\
        \bottomrule
    \end{tabular}
    \label{tab:example-model-refs}
\end{table}

\section{Evaluation} \label{section:evaluation}

This section describes all aspects of the empirical evaluation of the proposed algorithms. First, the selected subject programs used for all research questions are described. The considered evaluation metrics are then outlined, including the new metrics proposed by us. Finally, research questions are stated.

\subsection{Subject programs (RTPTorrent)}

The RTPTorrent dataset~\citep{mattis2020rtptorrent} is used for all evaluations. It consists of multiple open-source software projects. We consider this dataset to be the most suitable choice for the study, as the publication by \citet{mattis2020rtptorrent} is the only work identified in the SR that focuses on addressing issues that occur in other TCP datasets. Namely, RTPTorrent is characterized by the following qualities we deem desireable: it uses data from real software systems instead of synthetic cases; it includes a diverse range of Java repositories in terms of maturity, contributor count, test case count, and codebase size; it provides a non-trivial baseline based on a microstudy, which is a better reference point than base order or random order; it is publicly available and uses only open source projects from GitHub, which supports the reproducibility of research results; and it is compatible with the GHTorrent and TravisTorrent~\citep{beller2017travistorrent} datasets.

Arguably, the biggest issue with the dataset is that it only includes open-source Java repositories, which makes the results poorly generalizable to projects written in other programming languages and to proprietary systems. Furthermore, the granularity of the included test cases is low as test classes are used instead of test methods~\citep{mattis2020rtptorrent}.

Only the following RTPTorrent subject programs were excluded from the evaluation for the reasons outlined below:
\begin{itemize}
\item \texttt{Achilles}, \texttt{buck}, and \texttt{sonarqube} were excluded because their memory requirements exceeded the capacity of our evaluation environment. The testing framework loads a large number of resources simultaneously, making these projects too large to evaluate on our hardware.
\item \texttt{deeplearning4j}, \texttt{jOOQ}, \texttt{jcabi-github}, \texttt{graylog2-server}, and \texttt{jetty.project} were excluded due to a substantial number of missing cycles, commits, or files. In some cases, several hundred commits were missing associated files.
\item \texttt{sling} was excluded because its repository data is no longer available. RTPTorrent requires repositories to be cloned locally, and this repository became inaccessible during our research.
\end{itemize}
We include all 12 remaining projects of various sizes, which are listed in \Cref{tab:rtptorrent-repositories}.

\begin{table}[htbp]
    \centering
    \caption{The description of selected RTPTorrent repositories.}
    \resizebox{\textwidth}{!}{
    \begin{tabular}{l r r r r}
        \toprule
        \textbf{Subject program} & \textbf{\begin{tabular}[c]{@{}c@{}}Count of\\ CI cycles\end{tabular}} & \textbf{\begin{tabular}[c]{@{}c@{}}Failure\\ percentage\end{tabular}} & \textbf{\begin{tabular}[c]{@{}c@{}}Average number\\ of test cases\end{tabular}} & \textbf{\begin{tabular}[c]{@{}c@{}}Total time\\ spent in CI\end{tabular}} \\ \midrule
        \texttt{LittleProxy} & 427 & 6.6\% & 25.6 & 59\,h \\
        \texttt{HikariCP} & 1\,577 & 5.3\% & 14.1 & 76\,h \\
        \texttt{jade4j} & 931 & 9.1\% & 38.5 & 52\,h \\
        \texttt{wicket-bootstrap} & 904 & 2.7\% & 42.9 & 53\,h \\
        \texttt{titan} & 941 & 10.4\% & 41.7 & 1\,215\,h \\
        \texttt{dynjs} & 935 & 1.7\% & 73.3 & 229\,h \\
        \texttt{jsprit} & 1\,061 & 4.1\% & 86.5 & 137\,h \\
        \texttt{DSpace} & 1\,863 & 1.3\% & 62.5 & 326\,h \\
        \texttt{optiq} & 1\,306 & 1.8\% & 42.2 & 1\,011\,h \\
        \texttt{cloudify} & 4\,968 & 2.6\% & 55.1 & 1\,447\,h \\
        \texttt{okhttp} & 5\,425 & 6.5\% & 42.8 & 868\,h \\
        \bottomrule
    \end{tabular}
    }
    \label{tab:rtptorrent-repositories}
\end{table}

The included repositories should be sufficiently diverse, with cycle counts ranging from 427 to more than 5\,000, failed cycle percentages from 1.3\% to 10.4\%, and average test cases per cycle from 14.1 to more than 100. Finally, the total time spent during continuous integration jobs varies from a few dozen hours to a few months.

To enrich the dataset with the job execution time, the previously mentioned TravisTorrent~\citep{beller2017travistorrent} dataset was joined using the compatible job identifier. To include the source codes, the full version control system (VCS) repositories of the evaluated systems are downloaded, and the appropriate files are retrieved according to the commit identifiers given by RTPTorrent.

Some mismatch errors occur during the enrichment. That is, we encountered 4 errors for \texttt{LittleProxy}, 3 for \texttt{wicket-bootstrap}, 66 for \texttt{DSpace}, 5 for \texttt{cloudify}, and 72 for \texttt{okhttp}. We believe that the number of mismatches in selected projects is negligible and does not pose a threat to the validity of the empirical study, since in the worst case erroneous cycles constitute only a few percent of all job runs (for \texttt{DSpace}).

\subsection{Evaluation metrics}

All metrics presented in \Cref{section:related-work} were analyzed to select the right indicators for this study. NAPFD and the severity-aware version of \APFDc{} were ruled out, since we do not perform severity-aware or constrained prioritization. Then, RPA and NRPA were excluded because they treat all cycles equally (even though failed cycles are generally more important) and because sometimes well-performing permutations may receive low RPA/NRPA scores~\citep{bagherzadeh2022reinforcement,zhao2023revisiting}.

To rate the effectiveness of cost-unaware approaches, we prefer the usage of \rAPFD{} over APFD. The former has a consistent value range of exactly zero to one, where zero is achieved by the worst solution and one is achieved by the optimal ordering, yet has the same monotonicity as APFD~\citep{zhao2023revisiting}, making it safe to use. Unfortunately, no rectified version of \APFDc{} is presented in the analyzed literature.

The NTR metric is suitable for evaluating the applicability of a TCP method within CI environments. However, its interpretation might be misleading. Normalized time reduction indicates the percentage time savings, assuming that the CI cycle is stopped after the first failure. That said, these savings occur only for failed cycles, which constitute a tiny percentage of all cycles. Meanwhile, the overhead of prioritization must occur for all cycles. Thus, a metric including the unfavorable TCP overhead is sought after.

To evaluate the effectiveness of cost-aware (but severity-unaware) TCP methods, we propose a new metric called the \textit{rectified time-aware average percentage of faults detected} (\rAPFDc{}). Similarly to the \rAPFD{} metric introduced by \citet{zhao2023revisiting}, \rAPFDc{} is defined as a min-max normalization of a previously used indicator. The formula for \rAPFDc{} is shown in \Cref{eq:proposed-rapfdc}:
\begin{equation} \label{eq:proposed-rapfdc}
    r\text{APFD}_C\left(s\right) = \frac{\text{APFD}_C\left(s\right) - \text{APFD}_{C\,\min}\left(s\right)}{\text{APFD}_{C\,\max}\left(s\right) - \text{APFD}_{C\,\min}\left(s\right)}.
\end{equation}

The proposed metric inhibits the following desirable qualities: the values of \rAPFDc{} are bounded by 0 and 1 (\(0 \leq r\text{APFD}_C(s) \leq 1\)); the optimal solution always has \rAPFDc{} of 1 (\(r\text{APFD}_{C\,\max}(s) = 1\)); the worst solution always has \rAPFDc{} of 0 (\(r\text{APFD}_{C\,\min}(s) = 0\)); the metric has the same monotonicity as \APFDc{} -- if an ordering is better than another in terms of \rAPFDc{}, it is also better in terms of \APFDc{}.

\begin{figure}[htbp]
    \centering
    \includegraphics[width=0.4\textwidth]{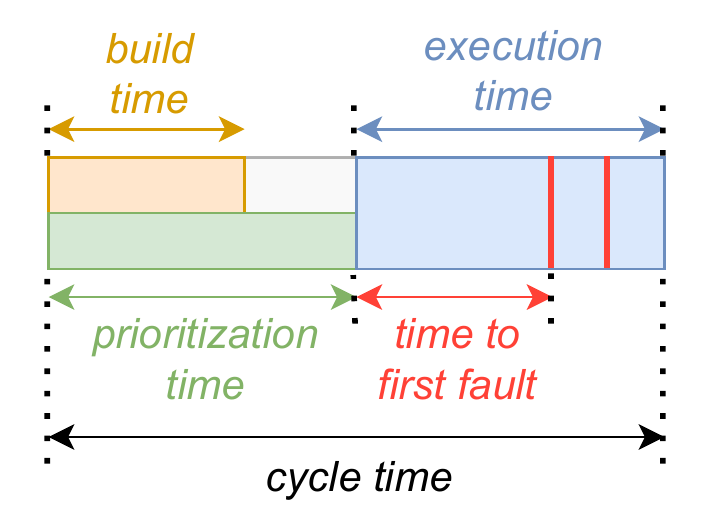}
    \caption{Diagram of various static TCP cycle time measures.}
    \label{fig:atr-measures}
\end{figure}

As stated previously, a new TCP CI applicability metric is needed for use alongside NTR. In \Cref{fig:atr-measures}, various durations within the CI regression testing are shown. If the TCP method used is static, we can perform prioritization in parallel with compilation and static analysis. If prioritization is faster than the build, there is no time cost to performing TCP; if it is slower, there exists an overhead.

Thus, we introduce \textit{actual time reduction} (ATR), which represents the genuine percentage of time savings from using a given TCP method in a given repository, as an alternative to the normalized time reduction (NTR) metric. The proposed ATR metric sums up the total testing time (TT) spent in CI jobs. The testing time is defined using the formula from \Cref{eq:proposed-testing-time}:
\begin{equation} \label{eq:proposed-testing-time}
    TT = \max\left(PT - BT,\,0\right) + \begin{cases}
        TF \quad & \text{if } TF \neq null \\
        EF \quad & \text{if } TF = null,
    \end{cases}
\end{equation}
\noindent where: prioritization time \(PT\) -- time spent performing TCP (or zero if TCP is not used); build time \(BT\) -- time spent on compilation and static analysis independently of testing; time to first fault \(TF\) -- duration until the end of the first failing test case in the sequence; execution time \(EF\) -- time spent on executing all test cases in the suite. Consequently, we define the actual time reduction (ATR) for approach \(\mathcal{A}\) as in \Cref{eq:proposed-atr}:
\begin{equation} \label{eq:proposed-atr}
    \text{ATR} = 1 - \dfrac{\sum\limits_{i=1}^{\text{CI}} TT_{\mathcal{A}}}{\sum\limits_{i=1}^{\text{CI}} TT_{\text{base}}}.
\end{equation}

All durations should be measured using common units. The intuitive interpretation of ATR is as follows: a value of 0.01 indicates that approach \(\mathcal{A}\) reduces the time spent on all testing activities by 1\%. Negative ATR indicates TCP methods that would be harmful to apply in practice. The newly introduced ATR metric is vastly different from NTR in terms of construction and interpretation. Thus, to avoid threats to the construct validity, questions regarding the applicability of approaches will be answered using both NTR and ATR.

\subsection{Research questions}

We aim to obtain answers to two high-level questions about the proposed work. Firstly, whether the approach combinators manage to beat the sub-approaches they combine. Secondly, whether the performance of proposed solutions is comparable with the current state of the art, and whether they can be used to perform TCP in real-world scenarios.

Consequently, the first research question (RQ) is stated as follows:\\
RQ1: \textit{Can approach combinators beat their sub-approaches in terms of effectiveness?} It is divided into multiple sub-questions. Mixers are covered in RQ1.1, interpolators in RQ1.2, and tiebreakers in RQ1.3. All approaches are compared in terms of \rAPFDc{} as cost-aware methods are also evaluated.

Since approach combinators form entire families of approaches, their performance can vary depending on the choice of the underlying base models. For example, if a mixer is configured with a high weight for a poorly performing method and a comparatively low weight for a strong method, the resulting prioritizer may be inferior to the stronger sub-approach. Consequently, evaluating RQ1 requires the selection of specific, concrete instances, with the understanding that the resulting findings do not generalize to all constructed combinators. To this end, we select the sample models from the previous section and compare them against all the sub-approaches that they employ. Accordingly, P1 is used for RQ1.1, P2 is used for RQ1.2, and P3 is used for RQ1.3.

The second research question investigates the performance of the proposed approach combinator examples (P1, P2, and P3) against solutions from the state of the art and has the following form:\\ 
RQ2: \textit{What is the performance of the proposed approaches?} The comparison includes two aspects, general effectiveness (measured using \rAPFDc{}) and applicability within CI (in terms of NTR and ATR).

The main difficulty related to RQ2 is the selection of representative comparison reference points, both in the form of trivial baselines and state-of-the-art approaches. We first gather all appropriate approaches and evaluate them in terms of \rAPFDc{}, and then select a~few of them as baselines to answer the question. Two major steps were used to find more comparison points: we checked the most popular baselines for TCP evaluation from \citet{pradolima2020test}, and collected all prioritization approaches of the previous five years found in the SR. A total of 60 techniques were reviewed, 54 were excluded for various reasons, and there were 3 duplicates. The complete overview of all considered algorithms can be found in Section 3 of our online Appendix to this paper\hyperref[fn:Appendix]{\textsuperscript{\ref*{fn:Appendix}}}.

In the end, three new techniques were included as baselines, ROCKET by \citet{marijan2013test}, DBP by \citet{zhou2021beating}, and an unnamed approach by \citet{fazlalizadeh2009incorporating}. This selection resulted from a process of elimination, as these were the only methods that did not meet the exclusion criteria mentioned in our Appendix. We justify the suitability of some of the selected baselines below:
\begin{itemize}
    \item Random prioritization is the most common reference point for TCP evaluation \citep{pradolima2020test}, and the base order holds second place ex aequo. They are both used in a multitude of related studies \citep{marijan2023comparative,fazlalizadeh2009incorporating,zhou2021beating,pradolima2022amultiarmed,spieker2017reinforcement,rothermel2001prioritizing,rothermel1999test,marijan2013test}.
    \item DFE~\citep{kim2002ahistorybased} is used as a non-trivial baseline in RTPTorrent~\citep{mattis2020rtptorrent}, which we build upon. It represents the use of simple TCP approaches.
    \item ROCKET~\citep{marijan2013test} is a robust approach, tied as the second most widely used TCP evaluation baseline \citep{pradolima2020test}. It uses information factors similar to the proposed methods and, despite being developed in 2013, was used as a reference point as recently as 2023~\citep{marijan2023comparative}. Most importantly, it has been shown to perform better than many ML methods~\citep{marijan2023comparative}. It represents the state-of-the-art heuristic TCP.
\end{itemize}

All stated RQs are described along with the used metrics in \Cref{tab:research-questions}.

\begin{table}[htbp]
    \centering
    \begin{tabular}{l l c}
    \toprule
    \textbf{Number} & \textbf{Question} & \textbf{Metrics} \\ \midrule
    RQ1: & \textit{\begin{tabular}[l]{@{}l@{}} Can approach combinators beat their \\ sub-approaches in terms of effectiveness? \end{tabular}} & --- \\
    \hspace{0.3cm}\textbullet\hspace{0.1cm} RQ1.1: & \textit{\begin{tabular}[l]{@{}l@{}} Can \textbf{mixers} beat their \\ sub-approaches in terms of effectiveness? \end{tabular}} & \rAPFDc{} \\
    \hspace{0.3cm}\textbullet\hspace{0.1cm} RQ1.2: & \textit{\begin{tabular}[l]{@{}l@{}} Can \textbf{interpolators} beat their \\ sub-approaches in terms of effectiveness? \end{tabular}} & \rAPFDc{} \\
    \hspace{0.3cm}\textbullet\hspace{0.1cm} RQ1.3: & \textit{\begin{tabular}[l]{@{}l@{}} Can \textbf{tiebreakers} beat their \\ base approaches in terms of effectiveness? \end{tabular}} & \rAPFDc{} \\ \midrule
    RQ2: & \textit{\begin{tabular}[l]{@{}l@{}} What is the performance \\ of the proposed approaches? \end{tabular}} & --- \\
    \hspace{0.3cm}\textbullet\hspace{0.1cm} RQ2.1: & \textit{\begin{tabular}[l]{@{}l@{}} What is the \textbf{effectiveness} \\ of the proposed approaches? \end{tabular}} & \rAPFDc{} \\
    \hspace{0.3cm}\textbullet\hspace{0.1cm} RQ2.2: & \textit{\begin{tabular}[l]{@{}l@{}} What is the \textbf{CI applicability} \\ of the proposed approaches? \end{tabular}} & NTR, ATR \\ \bottomrule
    \end{tabular}
    \caption{The summary of stated research questions.}
    \label{tab:research-questions}
\end{table}

\section{Results} \label{section:results}

All experiments were conducted on a 4-core 2.42 GHz CPU from 2022 and 16 GB of RAM. These hardware specifications can only affect the ATR metric, with no influence on other measures. To avoid misleading evaluation results, we count the APFD family metrics only for failed cycles, with no data imputation for others, and include only cycles with at least 6 test cases in the suite, as previously suggested by \citet{bagherzadeh2022reinforcement}.

Two forms of presentation of the results are used. Firstly, the average metric values are organized into tables. In the footers, the means and medians across all datasets are presented. Since all calculated metrics are normalized, it is safe to aggregate them in this way. For each row, the best result is bolded. Secondly, the results are visualized as box plots, showing the metric value distributions across all datasets. There, boxes show the range between the first and third quartiles, whiskers extend from the box to 1.5x the interquartile range, outliers are shown as dots, the median is presented as a horizontal line, and the mean is marked by a small triangle.

\subsection{Approach combinators compared to base methods (RQ1)}

\begin{figure}[htbp]
    \centering
    \includegraphics[width=0.49\textwidth]{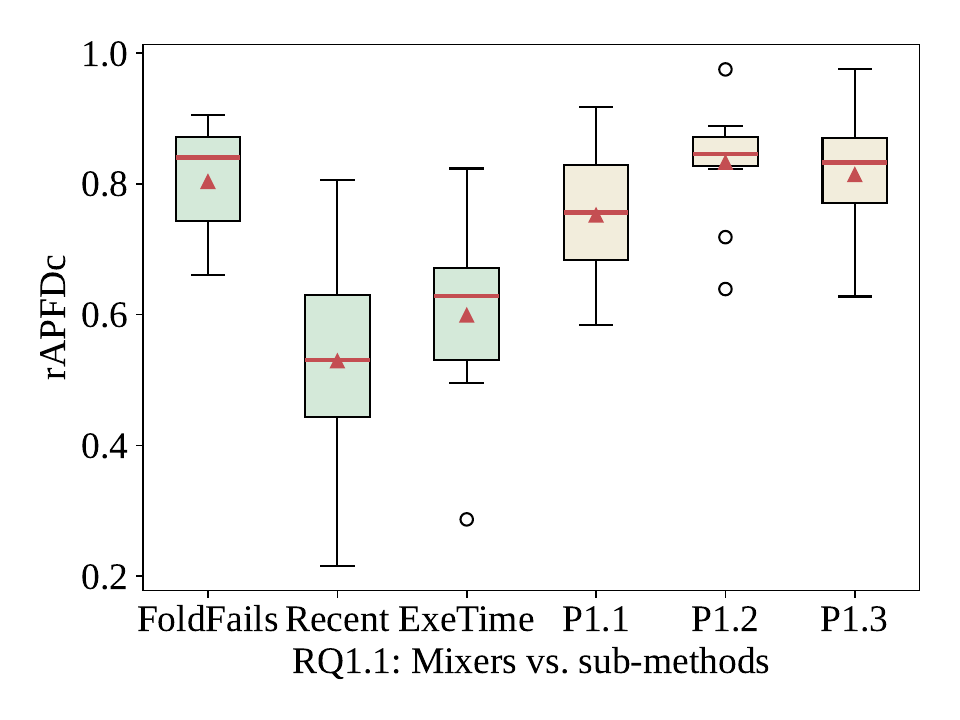}
    \hfill
    \includegraphics[width=0.49\textwidth]{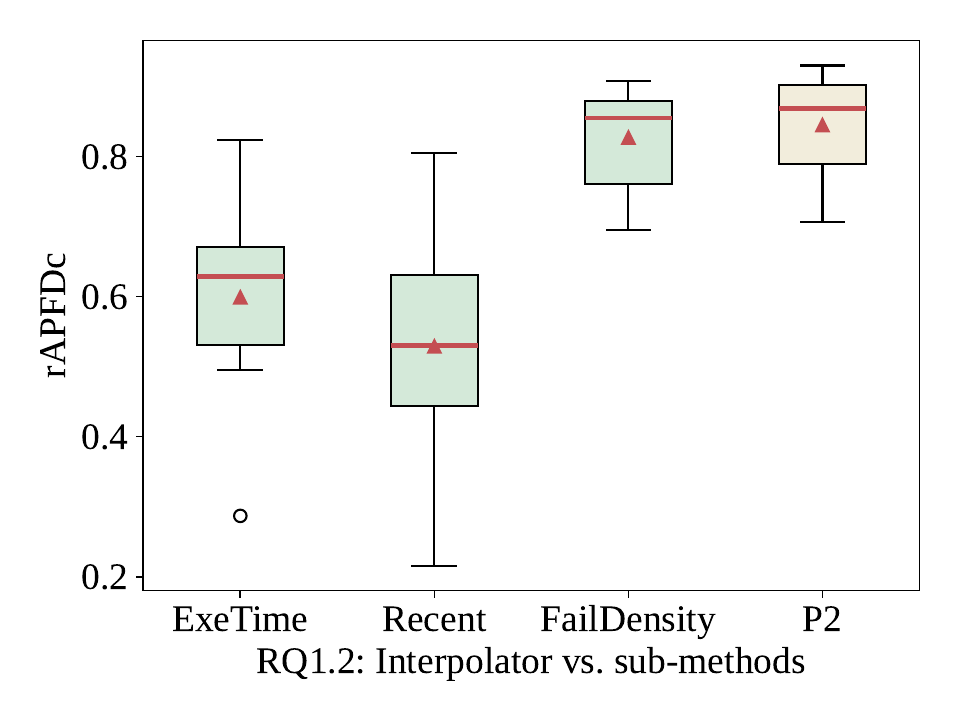}
    \caption{Boxplot of average \rAPFDc{} values for RQ1.1 and RQ1.2.}
    \label{fig:results-rq11-rq12}
\end{figure}

\begin{figure}[htbp]
    \centering
    \includegraphics[width=0.49\textwidth]{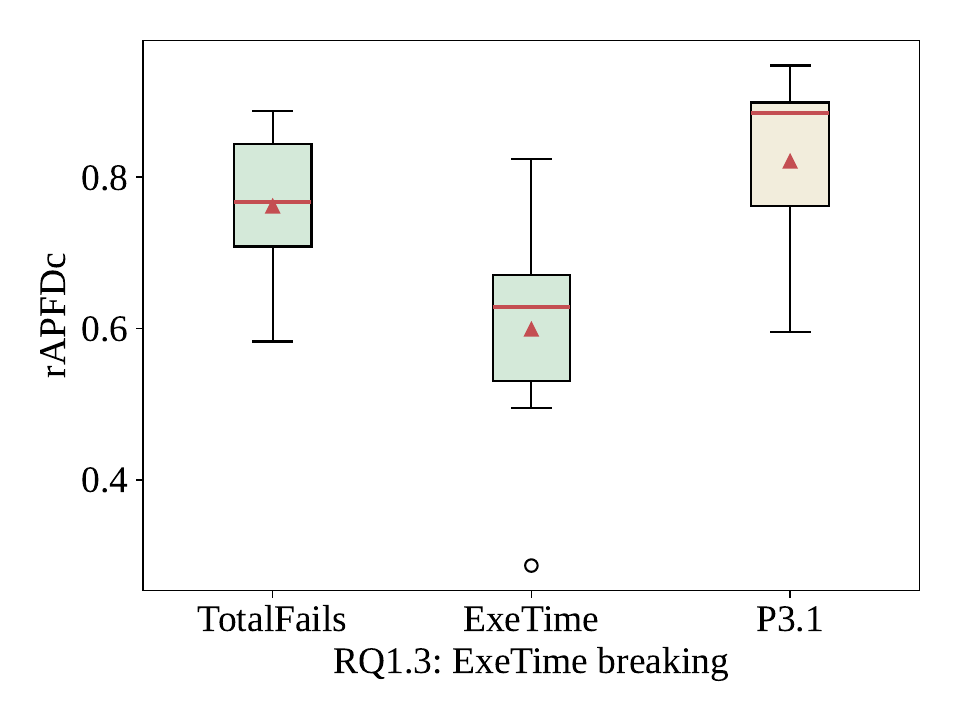}
    \hfill
    \includegraphics[width=0.49\textwidth]{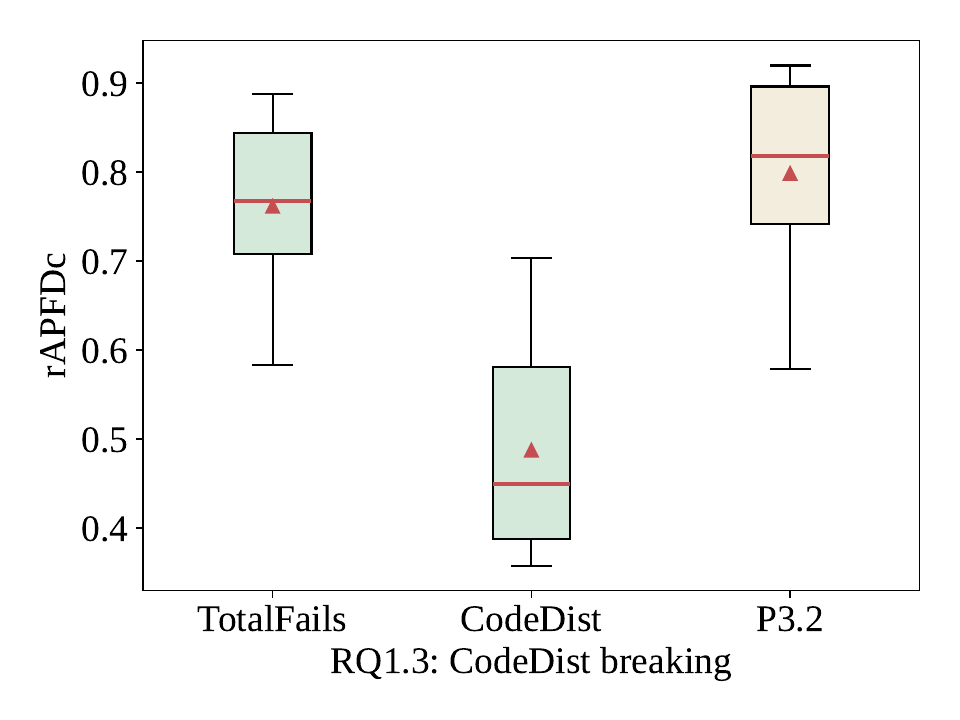}
    \caption{Boxplot of average \rAPFDc{} values for RQ1.3.}
    \label{fig:results-rq13}
\end{figure}

The evaluation metric values for RQ1.1 and RQ1.2 are presented in \Cref{tab:results-rq11-rq12} and \Cref{fig:results-rq11-rq12}, while the results for RQ1.3 are shown in \Cref{tab:results-rq13} and \Cref{fig:results-rq13}.

In RQ1.1, we compare all variants of the P1 approach (P1.1 -- random, P1.2 -- Borda, and P1.3 -- Schulze) against their sub-methods. We can observe the Borda approach achieving the highest results on 6 out of 11 subject programs (with three joint first places tied with Schulze). However, one of the sub-approaches, FoldFailsOrder, achieved the best effectiveness in the five remaining systems. The Borda mixer (P1.2) also has the highest mean and median \rAPFDc{} values.

For RQ1.2, the interpolator achieves the highest \rAPFDc{} metric value on 10 out of 11 datasets, with its best sub-method, FailDensityOrder, being better only for \texttt{dynjs}. It also has the highest mean and median \rAPFDc{}.

\begin{table}[htbp]
    \centering
    \caption{Average \rAPFDc{} per dataset for RQ1.1 and RQ1.2.}
    \resizebox{\textwidth}{!}{
\begin{tabular}{l r r r r r r}
    \toprule
    \multirow{2}{*}{\textbf{Subject program}} & \multicolumn{6}{c}{\textbf{RQ1.1: Mixers vs. sub-methods}} \\
     & \textbf{FoldFails} & \textbf{Recent} & \textbf{ExeTime} & \textbf{P1.1} & \textbf{P1.2} & \textbf{P1.3} \\ \midrule
    \texttt{LittleProxy} & \textbf{.697} & .375 & .501 & .584 & .639 & .628 \\
    \texttt{HikariCP} & .749 & .530 & .666 & .719 & \textbf{.846} & .831 \\
    \texttt{jade4j} & .661 & .805 & .657 & .652 & \textbf{.857} & .681 \\
    \texttt{wicket-bootstrap} & .748 & .727 & .495 & .756 & \textbf{.833} & \textbf{.833} \\
    \texttt{titan} & \textbf{.896} & .431 & .287 & .844 & .856 & .856 \\
    \texttt{dynjs} & .840 & .533 & .824 & .857 & \textbf{.888} & .885 \\
    \texttt{jsprit} & .872 & .623 & .601 & .917 & \textbf{.975} & \textbf{.975} \\
    \texttt{DSpace} & \textbf{.738} & .482 & .676 & .619 & .719 & .719 \\
    \texttt{optiq} & .870 & .216 & .692 & .814 & \textbf{.889} & \textbf{.889} \\
    \texttt{cloudify} & \textbf{.856} & .639 & .628 & .794 & .823 & .822 \\
    \texttt{okhttp} & \textbf{.905} & .456 & .560 & .715 & .834 & .835 \\
    \midrule
    \textbf{Mean} & .803 & .529 & .599 & .752 & \textbf{.833} & .814\\
    \textbf{Median} & .840 & .530 & .628 & .756 & \textbf{.846} & .833\\
    \bottomrule
\end{tabular}
    \hspace{0.5cm}
\begin{tabular}{r r r r}
    \toprule
    \multicolumn{4}{c}{\textbf{RQ1.2: Interpolator vs. sub-methods}} \\
    \textbf{ExeTime} & \textbf{Recent} & \textbf{FailDensity} & \textbf{P2} \\ \midrule
    .501 & .375 & .694 & \textbf{.707} \\
    .666 & .530 & \textbf{.775} & \textbf{.775} \\
    .657 & .805 & \textbf{.887} & \textbf{.887} \\
    .495 & .727 & .747 & \textbf{.785} \\
    .287 & .431 & .908 & \textbf{.930} \\
    .824 & .533 & \textbf{.846} & .831 \\
    .601 & .623 & .872 & \textbf{.896} \\
    .676 & .482 & .738 & \textbf{.793} \\
    .692 & .216 & .870 & \textbf{.914} \\
    .628 & .639 & .855 & \textbf{.868} \\
    .560 & .456 & .903 & \textbf{.908} \\
    \midrule
    .599 & .529 & .827 & \textbf{.845}\\
    .628 & .530 & .855 & \textbf{.868}\\
    \bottomrule
\end{tabular}
    }
    \label{tab:results-rq11-rq12}
\end{table}

As stated previously, for RQ1.3, two tiebreaker experiments were performed and their results are shown in \Cref{tab:results-rq13} and \Cref{fig:results-rq13}. The left half of both the table and the figure shows the evaluation of P3.1 (execution time tiebreaking), and the right half presents P3.2 (code representation distance tiebreaking). In the first half, P3.1 achieved the highest results for all subjects except \texttt{HikariCP} and \texttt{jade4j}. The mean and median \rAPFDc{} of the tiebreaker was also much higher than that of its sub-approaches. In the second half, P3.2 outperformed all other approaches in 10 out of 11 repositories, with the only exception being \texttt{HikariCP}. It also had the highest mean and median \rAPFDc{}, compared to its sub-methods. However, it performed worse than the execution time tiebreaker.

\begin{table}[htbp]
    \centering
    \caption{Average \rAPFDc{} per dataset for RQ1.3.}
    \resizebox{0.9\textwidth}{!}{
\begin{tabular}{l r r r}
    \toprule
    \multirow{2}{*}{\textbf{Subject program}} & \multicolumn{3}{c}{\textbf{RQ1.3: ExeTime breaking}} \\
     & \textbf{TotalFails} & \textbf{ExeTime} & \textbf{P3.1} \\ \midrule
    \texttt{LittleProxy} & .678 & .501 & \textbf{.716} \\
    \texttt{HikariCP} & .583 & \textbf{.666} & .596 \\
    \texttt{jade4j} & .606 & \textbf{.657} & .617 \\
    \texttt{wicket-bootstrap} & .744 & .495 & \textbf{.898} \\
    \texttt{titan} & .888 & .287 & \textbf{.899} \\
    \texttt{dynjs} & .835 & .824 & \textbf{.947} \\
    \texttt{jsprit} & .819 & .601 & \textbf{.898} \\
    \texttt{DSpace} & .739 & .676 & \textbf{.822} \\
    \texttt{optiq} & .852 & .692 & \textbf{.943} \\
    \texttt{cloudify} & .767 & .628 & \textbf{.809} \\
    \texttt{okhttp} & .865 & .560 & \textbf{.884} \\
    \midrule
    \textbf{Mean} & .761 & .599 & \textbf{.821}\\
    \textbf{Median} & .767 & .628 & \textbf{.884}\\
    \bottomrule
\end{tabular}
    \hspace{0.5cm}
\begin{tabular}{r r r}
    \toprule
    \multicolumn{3}{c}{\textbf{RQ1.3: CodeDist breaking}} \\
    \textbf{TotalFails} & \textbf{CodeDist} & \textbf{P3.2} \\ \midrule
    .678 & .614 & \textbf{.679} \\
    \textbf{.583} & .450 & .579 \\
    .606 & .379 & \textbf{.615} \\
    .744 & .557 & \textbf{.811} \\
    .888 & .505 & \textbf{.915} \\
    .835 & .397 & \textbf{.920} \\
    .819 & .604 & \textbf{.847} \\
    .739 & .427 & \textbf{.818} \\
    .852 & .703 & \textbf{.915} \\
    .767 & .358 & \textbf{.804} \\
    .865 & .368 & \textbf{.878} \\
    \midrule
    .761 & .488 & \textbf{.798}\\
    .767 & .450 & \textbf{.818}\\
    \bottomrule
\end{tabular}
    }
    \label{tab:results-rq13}
\end{table}

To summarize the entire research question, the approach of combinator-based models can effectively beat their sub-approaches in terms of \rAPFDc{}. Without any fine-tuning, we were able to construct sample approach combinator methods (P1, P2, and P3), which consistently outperform their base approaches across the majority of subject programs. Between all mixers tested, the one that used Borda count (P1.2) had the best effectiveness, and in terms of RQ1.3, the execution time tiebreaker (P3.1) was better than its alternative. Thus, for the comparison in RQ2, the P1.2, P2, and P3.1 prioritizers will be used. As noted when first stating the research question, these findings cannot be generalized to all constructible combinator models, as the performance of such combinators depends on their specific configuration. Accordingly, the observed performance characteristics apply only to the specific sample prioritizers evaluated.

\subsection{Performance of the proposed approaches (RQ2)}

\begin{figure}[htbp]
    \centering
    \includegraphics[width=0.49\textwidth]{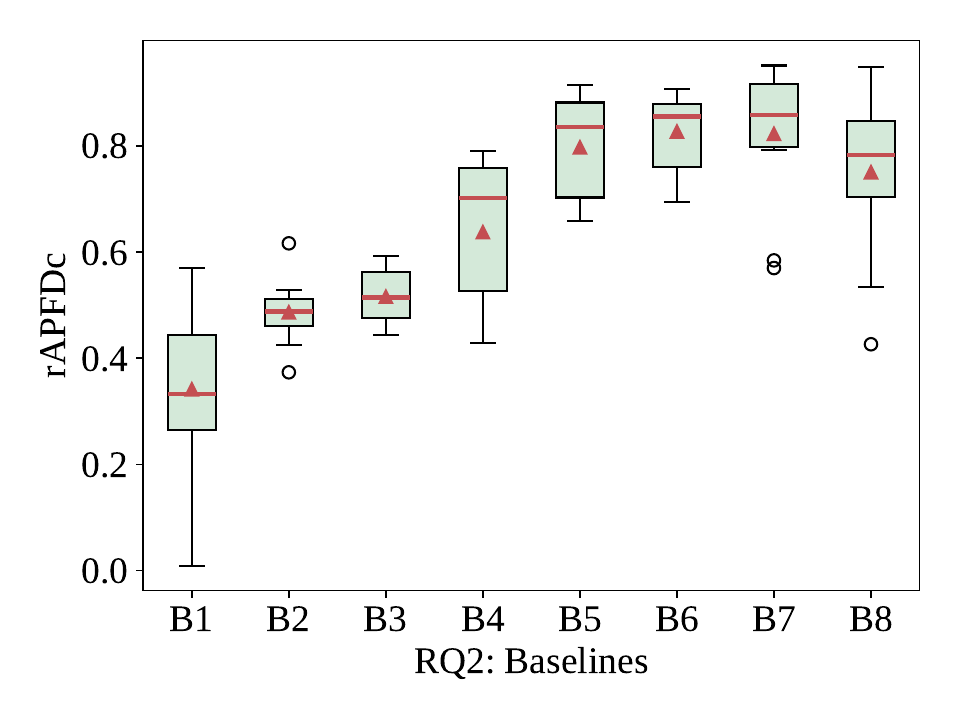}
    \hfill
    \includegraphics[width=0.49\textwidth]{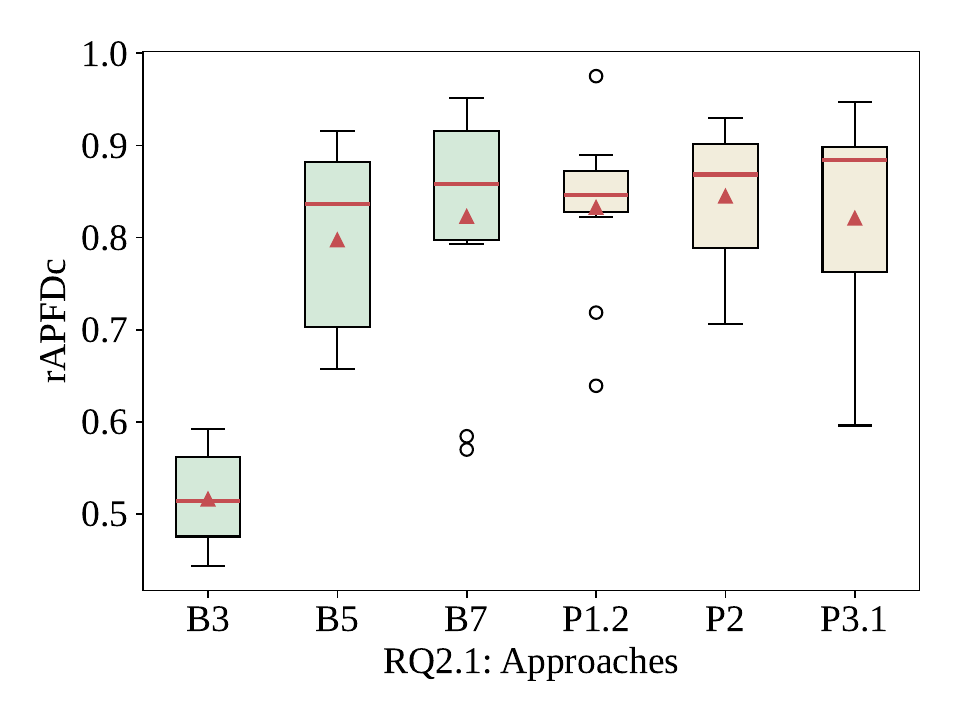}
    \caption{Boxplot of average \rAPFDc{} values for baselines and RQ2.1.}
    \label{fig:results-rq21}
\end{figure}

\begin{figure}[htbp]
    \centering
    \includegraphics[width=0.49\textwidth]{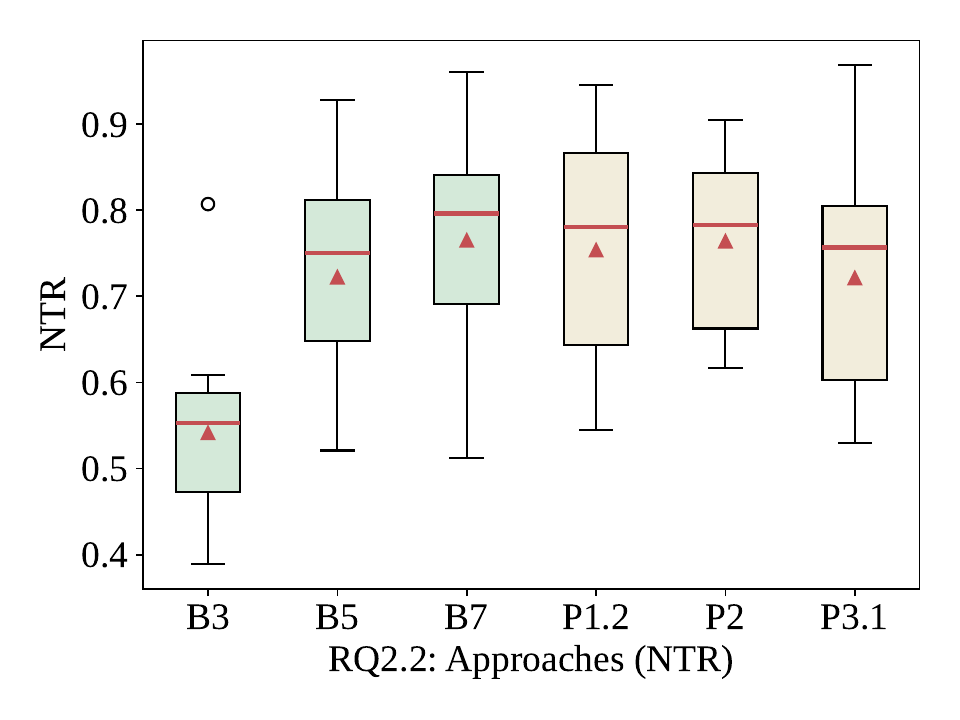}
    \hfill
    \includegraphics[width=0.49\textwidth]{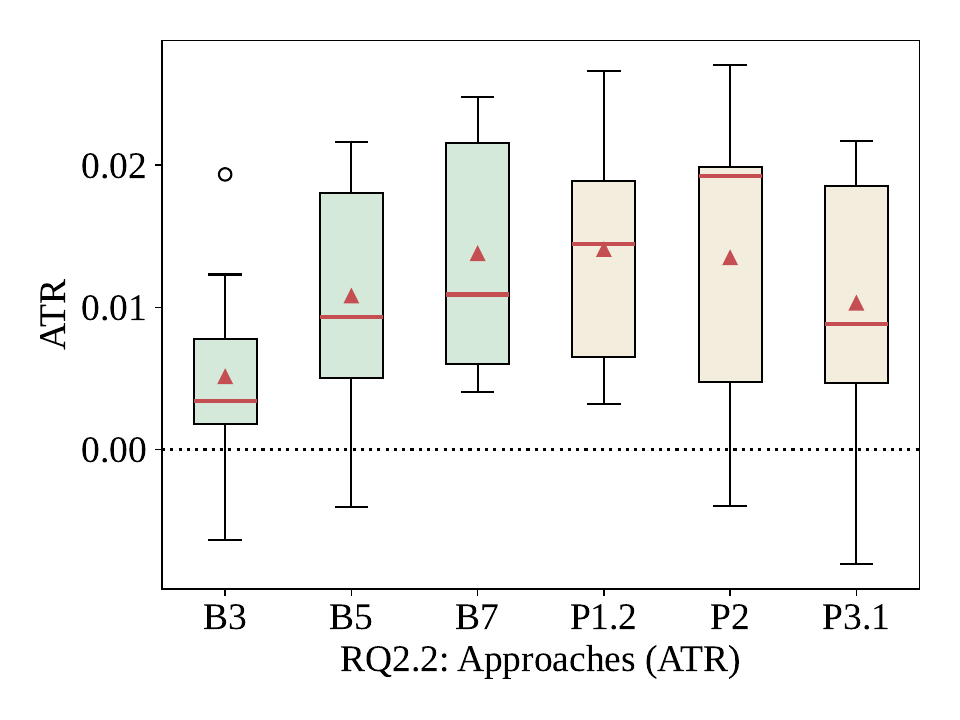}
    \caption{Boxplot of average NTR and ATR values for RQ2.2.}
    \label{fig:results-rq22}
\end{figure}

The performance of the proposed approaches is compared with both trivial baselines and state-of-the-art solutions, in terms of both effectiveness (\rAPFDc{}; \Cref{tab:results-rq21} and \Cref{fig:results-rq21}) and applicability within CI (NTR and ATR; \Cref{tab:results-rq22} and \Cref{fig:results-rq22}).

First, to select suitable reference points for further comparison, a preliminary analysis of \rAPFDc{} of all baselines is performed. We compare: BaseOrder~\citep{rothermel1999test}~(B1), DBP~\citep{zhou2021beating}~(B2), RandomOrder~\citep{rothermel1999test}~(B3), unnamed approach by \citet{fazlalizadeh2009incorporating}~(B4), DFE~\citep{kim2002ahistorybased}~(B5), FailDensityOrder~(B6), and ROCKET~\citep{marijan2013test} with \(m = 100\) (B7) and \(m = 1000\) (B8). The results of this microstudy are shown in the left section of \Cref{tab:results-rq21} and \Cref{fig:results-rq21}. We select B3, which is the random order of the test cases, as the lowest baseline. From B4 through B6, we select DFE, as it was previously used as a reference point by \citet{mattis2020rtptorrent}. Finally, we select B7 as the better-performing variant of ROCKET. In summary, for further evaluation, we use the following baselines: random prioritization (B3), DFE (B5), and ROCKET (B7).

We collect the \rAPFDc{} results for B3, B5, B7, P1.2, P2, and P3.1, which are shown in \Cref{tab:results-rq21} and \Cref{fig:results-rq21}. The best approaches vary by dataset, with P2 having the highest mean and P3.1 the highest median \rAPFDc{}. Random prioritization and DFE do not achieve the highest result for any dataset. It seems that all proposed approaches can compete with ROCKET in terms of effectiveness. ROCKET also outperforms P1.2 in terms of mean \rAPFDc{} and P3.1 in terms of average.

\begin{table}[htbp]
    \centering
    \caption{Average \rAPFDc{} per dataset for baselines and RQ2.1.}
    \resizebox{\textwidth}{!}{
\begin{tabular}{l r r r r r r r r}
    \toprule
    \multirow{3}{*}{\textbf{Subject program}} & \multicolumn{8}{c}{\textbf{RQ2: Baselines}} \\
    & \textbf{B1} & \textbf{B2} & \textbf{B3} & \textbf{B4} & \textbf{B5} & \textbf{B6} & \textbf{B7} & \textbf{B8} \\
     & \citep{rothermel1999test} & \citep{zhou2021beating} & \citep{rothermel1999test} & \citep{fazlalizadeh2009incorporating} & \citep{kim2002ahistorybased} & --- & \multicolumn{2}{c}{\citep{marijan2013test}} \\ \midrule 
    \texttt{LittleProxy} & .453 & .424 & .558 & .470 & .659 & \textbf{.694} & .584 & .426 \\
    \texttt{HikariCP} & .570 & .462 & .446 & .527 & .744 & .775 & \textbf{.793} & .535 \\
    \texttt{jade4j} & .226 & .485 & .566 & .769 & .661 & \textbf{.887} & .570 & .842 \\
    \texttt{wicket-bootstrap} & .009 & .529 & .514 & .741 & \textbf{.892} & .747 & .855 & .783 \\
    \texttt{titan} & .492 & .488 & .520 & .524 & .916 & .908 & \textbf{.930} & .719 \\
    \texttt{dynjs} & .398 & .514 & .494 & .791 & .871 & .846 & \textbf{.951} & .948 \\
    \texttt{jsprit} & .433 & .373 & .573 & .758 & .913 & .872 & \textbf{.936} & .724 \\
    \texttt{DSpace} & .303 & .616 & .592 & .702 & .658 & .738 & .801 & \textbf{.851} \\
    \texttt{optiq} & .205 & .508 & .457 & .429 & .836 & \textbf{.870} & .858 & .840 \\
    \texttt{cloudify} & .333 & .492 & .443 & .758 & .849 & .855 & \textbf{.902} & .898 \\
    \texttt{okhttp} & .333 & .459 & .512 & .544 & .771 & \textbf{.903} & .869 & .687 \\
    \midrule
    \textbf{Mean} & .341 & .486 & .516 & .638 & .797 & \textbf{.827} & .823 & .750\\
    \textbf{Median} & .333 & .488 & .514 & .702 & .836 & .855 & \textbf{.858} & .783\\
    \bottomrule
\end{tabular}
    \hspace{0.5cm}
\begin{tabular}{r r r r r r}
    \toprule
    \multicolumn{6}{c}{\textbf{RQ2.1: Approaches}} \\
    \textbf{B3} & \textbf{B5} & \textbf{B7} & \textbf{P1.2} & \textbf{P2} & \textbf{P3.1} \\
    \citep{rothermel1999test} & \citep{kim2002ahistorybased} & \citep{marijan2013test} & \multicolumn{3}{c}{\textbf{ours}} \\ \midrule
    .558 & .659 & .584 & .639 & .707 & \textbf{.716} \\
    .446 & .744 & .793 & \textbf{.846} & .775 & .596 \\
    .566 & .661 & .570 & .857 & \textbf{.887} & .617 \\
    .514 & .892 & .855 & .833 & .785 & \textbf{.898} \\
    .520 & .916 & \textbf{.930} & .856 & \textbf{.930} & .899 \\
    .494 & .871 & \textbf{.951} & .888 & .831 & .947 \\
    .573 & .913 & .936 & \textbf{.975} & .896 & .898 \\
    .592 & .658 & .801 & .719 & .793 & \textbf{.822} \\
    .457 & .836 & .858 & .889 & .914 & \textbf{.943} \\
    .443 & .849 & \textbf{.902} & .823 & .868 & .809 \\
    .512 & .771 & .869 & .834 & \textbf{.908} & .884 \\
    \midrule
    .516 & .797 & .823 & .833 & \textbf{.845} & .821\\
    .514 & .836 & .858 & .846 & .868 & \textbf{.884}\\
    \bottomrule
\end{tabular}
    }
    \label{tab:results-rq21}
\end{table}

As mentioned previously, we evaluate the applicability within CI in terms of both NTR and ATR to reduce the risks caused by the use of a new evaluation metric. The results are shown in \Cref{tab:results-rq22} and \Cref{fig:results-rq22}. While NTR has the standard value range of zero to one, ATR is technically not bounded and can take both positive and negative values.

In terms of NTR, the ROCKET approach has the highest mean and median metric value; however, its results are close to those of the proposed approaches. In particular, for P2, which performed the best, the mean NTR is lower by less than one percent. Although the best methods vary between subject programs, neither random prioritization nor DFE achieves the highest results for any repository.

\begin{table}[htbp]
    \centering
    \caption{Average NTR and ATR per dataset for RQ2.2.}
    \resizebox{\textwidth}{!}{
\begin{tabular}{l r r r r r r}
    \toprule
    \multirow{3}{*}{\textbf{Subject program}} & \multicolumn{6}{c}{\textbf{RQ2.2: Approaches (NTR)}} \\
    & \textbf{B3} & \textbf{B5} & \textbf{B7} & \textbf{P1.2} & \textbf{P2} & \textbf{P3.1} \\
    & \citep{rothermel1999test} & \citep{kim2002ahistorybased} & \citep{marijan2013test} & \multicolumn{3}{c}{\textbf{ours}} \\ \midrule
    \texttt{LittleProxy} & .440 & .521 & .513 & .564 & \textbf{.617} & .570 \\
    \texttt{HikariCP} & .608 & .634 & .796 & \textbf{.835} & .635 & .590 \\
    \texttt{jade4j} & .807 & .750 & .863 & .900 & \textbf{.904} & .630 \\
    \texttt{wicket-bootstrap} & .472 & .807 & .768 & .747 & .746 & \textbf{.824} \\
    \texttt{titan} & .558 & .779 & \textbf{.825} & .703 & .816 & .757 \\
    \texttt{dynjs} & .590 & .928 & .961 & .945 & .870 & \textbf{.968} \\
    \texttt{jsprit} & .585 & .824 & .857 & \textbf{.899} & .783 & .785 \\
    \texttt{DSpace} & .553 & .556 & .631 & .544 & \textbf{.658} & .616 \\
    \texttt{optiq} & .474 & .817 & .693 & .781 & \textbf{.890} & .883 \\
    \texttt{cloudify} & .389 & .662 & \textbf{.690} & .583 & .667 & .529 \\
    \texttt{okhttp} & .479 & .666 & \textbf{.818} & .789 & .817 & .781 \\
    \midrule
    \textbf{Mean} & .541 & .722 & \textbf{.765} & .754 & .764 & .721\\
    \textbf{Median} & .553 & .750 & \textbf{.796} & .781 & .783 & .757\\
    \bottomrule
\end{tabular}
    \hspace{0.5cm}
\begin{tabular}{r r r r r r}
    \toprule
    \multicolumn{6}{c}{\textbf{RQ2.2: Approaches (ATR)}} \\
    \textbf{B3} & \textbf{B5} & \textbf{B7} & \textbf{P1.2} & \textbf{P2} & \textbf{P3.1} \\
    \citep{rothermel1999test} & \citep{kim2002ahistorybased} & \citep{marijan2013test} & \multicolumn{3}{c}{\textbf{ours}} \\ \midrule
    +.000 & +.009 & +.008 & +.014 & \textbf{+.020} & +.015 \\
    -.006 & -.004 & +.011 & \textbf{+.014} & -.004 & -.008 \\
    +.019 & +.015 & +.024 & \textbf{+.027} & \textbf{+.027} & +.005 \\
    +.012 & \textbf{+.021} & +.020 & +.020 & +.020 & \textbf{+.022} \\
    +.006 & +.022 & \textbf{+.025} & +.016 & \textbf{+.024} & +.020 \\
    +.001 & \textbf{+.005} & \textbf{+.006} & \textbf{+.006} & +.005 & \textbf{+.006} \\
    +.009 & +.021 & +.023 & \textbf{+.025} & +.019 & +.019 \\
    +.003 & +.003 & \textbf{+.004} & +.003 & \textbf{+.004} & \textbf{+.004} \\
    +.003 & \textbf{+.008} & +.006 & +.007 & \textbf{+.009} & \textbf{+.009} \\
    +.002 & \textbf{+.005} & \textbf{+.005} & \textbf{+.004} & \textbf{+.005} & +.003 \\
    +.005 & +.013 & \textbf{+.019} & +.018 & \textbf{+.019} & +.018 \\
    \midrule
    +.005 & +.011 & \textbf{+.014} & \textbf{+.014} & \textbf{+.013} & +.010\\
    +.003 & +.009 & +.011 & +.014 & \textbf{+.019} & +.009\\
    \bottomrule
\end{tabular}
    }
    \label{tab:results-rq22}
\end{table}

For ATR, P2 has the highest median value, while the mean value is tied between joint winners: ROCKET, P1.2, and P2. We observe higher ATR values for more sophisticated yet lightweight methods. The value of ATR for the base order is also interpretable, as it would be equal to exactly zero in all datasets. One subject program, \texttt{HikariCP}, had negative ATR values for the majority of prioritization approaches.

When interpreted as percentage values, ATR is actually low in all areas, and the best approaches reduce the prioritization time by only up to 2.7\%. However, since TCP cannot reduce the actual execution time of particular test cases and also incurs overhead of its own, the reduction is actually quite large. In total terms, the proposed approaches saved, respectively 7.5\,h, 11.1\,h, and 9.2\,h for \texttt{titan}; 5.7\,h, 6.8\,h, and 6.8\,h for \texttt{optiq}; and 2.9\,h, 3.1\,h, and 2.7\,h for \texttt{okhttp}. These time gains may translate into substantial financial savings for these projects.

Based on the results of RQ2.1 and RQ2.2, we conclude that the proposed approach combinator prioritizers achieve performance comparable to current heuristic state-of-the-art methods, while employing a novel technique that, unlike many ML-based approaches, requires no training and has a low memory footprint, making it well suited for smaller development organizations.

This trade-off makes approach combinators particularly advantageous in several scenarios: (1)~resource-constrained environments where CI/CD pipelines have limited computational capacity; (2)~cold-start situations where insufficient historical data prevents effective ML model training; (3)~contexts requiring transparency, where explainable prioritization decisions are necessary; and (4)~rapid prototyping, where practitioners need to quickly experiment with different sub-approach combinations without retraining models.

To support the claim of comparable performance, we performed a statistical test on the results of RQ2.1. The Friedman test, previously used in the field of TCP~\citep{marijan2023comparative,zhao2023revisiting}, is employed. It is a non-parametric test, resistant to non-normality and differing variances~\citep{demsar2006statistical}. The test calculates the average ranks of the approaches in all subject programs to check whether the performance differences are statistically significant. If the Friedman test succeeds at a given significance level \(\alpha\), we continue with the post hoc Wilcoxon test to determine which pairs have a significant difference~\citep{demsar2006statistical}. As multiple hypotheses are tested, the procedure must be adjusted, in this case with Holm's method.

\begin{figure}[htbp]
    \centering
    \includegraphics[width=0.75\textwidth]{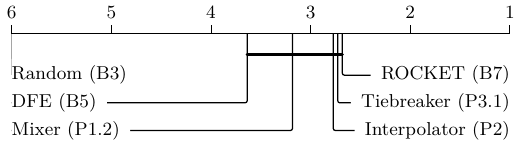}
    \caption{Critical difference plot for RQ2.1 \rAPFDc{} of different approaches.}
    \label{fig:rq2-cdd}
\end{figure}

For data visualization, we use critical difference (CD) diagrams, originally proposed by \citet{demsar2006statistical}. The CD plot shows the mean positions of the techniques, where the lower results are better and are displayed more to the right. Treatments are connected by a~horizontal line if there are no statistically significant differences between their results. In contrast, they are disconnected if significant differences occur. The results of the test are shown in \Cref{fig:rq2-cdd}. We can see that in terms of \rAPFDc{} performance, ROCKET wins but is closely followed by the proposed approaches P2 and P3.1. Then, the P1.2 and DFE approaches follow, and the random approach is the last. However, there are no statistically significant differences (at \(\alpha = 0.05\)) between the performance of all the state-of-the-art and proposed methods, as marked by the horizontal line.

\section{Discussion} \label{section:discussion}

The discussion chapter contains general insights from the research process. Threats to the validity of the results of the study are elaborately listed. Then, recommendations for other researchers, practitioners, and further work directions are outlined.

\subsection{Threats to validity}

\textit{Internal validity:} The main threat to internal validity concerns the correctness of the implementation of the proposed solution. Multiple tactics were used to reduce this risk. The entire contributed project uses typechecking and linting, with all rules enabled by default. The proposed framework has correctness precautions built in. Finally, the research artifacts are published for external inspection. In terms of SR, the number of involved researchers poses the primary threat to its internal validity. To address this threat, state-of-the-art guidelines~\citep{wohlin2014guidelines,moher2015preferred,kitchenham2023segress} were used, a SR protocol was written, and then approved by the second author. The use of the snowballing process should also reduce internal validity threats.

\textit{External validity:} The main risk of external validity comes from the fact that only open-source Java projects were tested. The reached conclusions might not generalize to projects written in other languages or to proprietary solutions. We believe that despite the reported threats, the use of RTPTorrent was the correct choice for the study. The dataset contains real-world data; it includes projects with diverse maturities, contributor counts, test case counts, codebase sizes, and failure percentages. For the SR, external validity is slightly concerned with the choice to select only peer-reviewed studies written in English. However, the vast majority of articles related to TCP are written in English. Furthermore, due to the use of Google Scholar, the external validity threats resulting from publisher and database bias are greatly reduced~\citep{wohlin2014guidelines}.

\textit{Construct validity:} The primary risks in this area refer to the use of less popular \rAPFD{} instead of APFD, and the use of newly proposed \rAPFDc{} and ATR metrics. The threat arising from the usage of \rAPFD{} and \rAPFDc{} is considered minimal. Most importantly, these metrics have the same monotonicity as APFD and \APFDc{}, respectively. Thus, it does not harm the relative comparisons of different algorithms. In terms of absolute evaluation, the rectified metrics have much more convenient semantics. For the proposed ATR metric, although it has a logical interpretation, it is very different from NTR. To eliminate associated risks, all applicability results are reported in terms of both metrics. Furthermore, we avoid invalid measurement practices reported in the literature~\citep{bagherzadeh2022reinforcement,zhao2023revisiting}.

\subsection{Recommendations for researchers and practitioners}

\textit{Validity of TCP research:} We share our insights to increase the validity of the TCP research ecosystem. Firstly, following \citet{mattis2020rtptorrent}, we observe that TCP studies are often performed on proprietary data or projects that do not reflect real-world software development practices. Secondly, not all studies are reproducible, and some do not publish their results at all~\citep{catal2013test}. Finally, incorrect use of metrics occurs in the literature~\citep{bagherzadeh2022reinforcement,zhao2023revisiting}. We suggest adhering to the following recommendations. Studies should use open-source datasets with diverse subject programs to support reproducible research and increase ecological validity. Their results should be compared with baselines better than random prioritization. Lastly, researchers should be careful when calculating evaluation metrics; in particular, we propose using rectified metrics from the APFD family and averaging over only failed cycles.

\textit{Unification of TCP:} We observe a lack of unification within the TCP research. We recommend that future researchers focus on contributing universal datasets and tools for TCP evaluation. The proposed TCPFramework is an attempt at such a~contribution. We also suggest the creation of a new comprehensive TCP approach taxonomy.

\textit{Approach combinators:} There are areas for improvement of the methodology. Firstly, additional combinators and categories may be proposed. Although combinators are applicable to almost all TCP methods, we only combined heuristic methods. Furthermore, we see an option to use supervised ML strategies on some combinators; namely, the weights of mixers may be automatically learned.
Based on our empirical observations, we offer guidelines for effective combinator construction: (1)~combine methods exploiting complementary information sources, as the best-performing combinators in our study integrated orthogonal factors; (2)~use balanced weights unless empirically justified otherwise; (3)~prefer theoretically grounded aggregation methods such as Borda count over random selection. Configurations combining redundant methods or using extreme weight distributions should be avoided.

\textit{Practical usage:} To sum up all the insights from both the SR and empirical experimentation, the primary guidance for practitioners is that TCP can provide measurable gains for CI processes, even when extremely simple methods are used. Simultaneously understanding the difficulties of integrating heavy ML-based TCP solutions in CI pipelines, we recommend the use of lightweight prioritization methods such as ones constructed with the approach combinator framework, which do not require training or historical testing data. We endorse using TCP methods that incorporate different information factors, as such techniques achieved higher performance in our empirical evaluations. The use of approach combinators allows for mixing these factors, without implementing specialized hybrid algorithms manually.

\section{Conclusions}\label{sec:Conclusions}

The literature survey conducted led to a few key insights. There are many TCP evaluation tools available, and none of them are universally used. The choice of the dataset should be influenced by the exact problem statement and other factors. For general use, we recommend the RTPTorrent~\citep{mattis2020rtptorrent} dataset, since it is publicly available and includes diverse subject programs. It is hard to establish the state-of-the-art TCP algorithms as studies from this area are mostly incomparable due to divergences in datasets, metrics, tools, information factors, and problem statements used. Some studies do not empirically evaluate their proposed approaches at all. The most widely used evaluation reference points are base order, random order, and the ROCKET~\citep{marijan2013test} algorithm.

Similarly, the empirical study uncovered answers to the questions regarding the applicability of approach combinators. Prioritizers based on the approach combinators framework can consistently outperform their sub-approaches in the evaluated examples. We recommend using approach combinators to merge the insights from prioritizers facilitating different information factors. The proposed methods based on approach combinators achieve performance comparable to the current state of the art, showing no statistically significant differences in terms of relative \rAPFDc{} ranking, while using a distinct technique. In terms of applicability, all proposed approaches consistently reduced the time spent testing in all but one subject system.

Although many further work opportunities remain, this work marks a step forward in addressing the problem of constructing universal, efficient, and applicable test case prioritization approaches, to reduce the costs of software testing.

\section*{CRediT authorship contribution statement}
\textbf{Tomasz Chojnacki}: Conceptualization, Methodology, Software, Validation, Investigation, Writing -- Original Draft, Writing -- Review \& Editing, Visualization.\\
\textbf{Lech Madeyski}: Conceptualization, Validation, Resources, Writing -- Review \& Editing, Supervision, Project administration.

\section*{Data Availability}

Reproducible research (RR) can address problems with the validity of findings in software engineering research~\citep{madeyski2017would}. To support RR and good scientific practices, the research is backed by project files in the form of code, scripts, and SR artifacts. The complete source code and instructions for obtaining and preparing the dataset are available online\hyperref[fn:Reproduction]{\textsuperscript{\ref*{fn:Reproduction}}}.
See also Section 2 of our online Appendix for details on running the code and extending the project\hyperref[fn:Appendix]{\textsuperscript{\ref*{fn:Appendix}}}.

\section*{Declaration of competing interest} 

The authors declare that they have no known competing financial interests or personal relationships that could have appeared to influence the work reported in this paper.

\end{document}